\documentclass[a4paper,fleqn,usenatbib]{mnras}

\usepackage{newtxtext,newtxmath,deluxetable}
\usepackage[T1]{fontenc}
\usepackage{ae,aecompl}

\usepackage{graphicx}	% Including figure files
\usepackage{amsmath}	% Advanced maths commands
\usepackage{amssymb}	% Extra maths symbols

\title[The High-Energy Rollover in 1H 0419-577]{On the Nature of the High-Energy Rollover in 1H 0419-577}

\author[T.J.Turner et al.]{
T.J.Turner,$^{1,2}$\thanks{E-mail: tjturner@umbc.edu}
J.N.Reeves,$^{2,3}$
V.Braito$^{2,4}$
and M Costa$^{3}$
\\
$^{1}$Department of Physics, University of Maryland Baltimore County, 
   Baltimore, MD 21250 U.S.A\\
$^{2}$Center for Space Science and Technology, University of Maryland Baltimore County, 1000 Hilltop Circle, Baltimore, MD 21250, USA\\
$^{3}$Astrophysics Group, School of Physical and Geographical Sciences, Keele 
University, Keele, Staffordshire ST5 5BG, U.K\\
$^{4}$INAF - Osservatorio Astronomico di Brera, Via Bianchi 46 I-23807 Merate (LC), Italy\\
}

\date{Accepted XXX. Received YYY; in original form ZZZ}

\pubyear{2017}

\begin{document}
\label{firstpage}
\pagerange{\pageref{firstpage}--\pageref{lastpage}}
\maketitle

\begin{abstract}
A {\it NuSTAR}/{\it Swift}  observation of the luminous Seyfert 1 galaxy 1H 0419-577 taken during 2015 reveals one of the most extreme high energy cut-offs observed to date from an AGN - an origin due to thermal Comptonization would imply a remarkably low coronal temperature $kT \sim 15$ keV.  The low energy peak of the spectrum in the hard X-ray 
{\it NuSTAR} band, which peaks before the expected onset of a Compton hump, rules out strong reflection as the origin of the hard excess in this AGN. 
%Indeed lack of a 
%strong Fe K$\alpha$ emission line, coupled with a weak Compton hump above 10 keV in the {\it NuSTAR} data show that reflection off the inner accretion disc %cannot explain the broadband spectral data obtained during 2015. 
We show the origin of the high energy rollover is likely due to a combination of both thermal Comptonization and a intrinsically steeper continuum, which is modified by absorption at lower energies. Furthermore, modeling the broadband XUV continuum shape as a colour-corrected accretion disc, requires the presence of a variable warm absorber to explain all flux and spectral states of the source, consistent with the previous work on this AGN. While absorber variations produce marked spectral variability in this AGN, consideration of all flux states allows us to isolate a colourless component of variability that may arise from changes in the inner accretion flow, typically at around $10 \, r_g$. 

\end{abstract}

\begin{keywords}
galaxies:active -- galaxies: individual: 1H 0419-577 -- galaxies: Seyfert -- X-rays: galaxies
\end{keywords}

%%%%%%%%%%%%%%%%%%%%%%%%%%%%%%%%%%%%%%%%%%%%%%%%%%

%%%%%%%%%%%%%%%%% BODY OF PAPER %%%%%%%%%%%%%%%%%%

\section{Introduction}

Observations with the {\it Ginga} satellite \citep[e.g.][]{pounds90a}  first measured a strong X-ray spectral 
component above 10 keV in type I active galactic nuclei (AGN),   initially interpreted entirely as  
reflection of hard X-rays off cold, Compton thick matter in the form of an accretion disc.  

However, X-ray measurements above 10 keV have been challenging both spatially and spectrally, leaving the detailed 
properties, and hence the origin of the hard X-ray flux open to debate. 
Scenarios involving partial covering of the continuum by Compton-thick matter \citep[e.g.][]{turner09b} or  Comptonization of soft seed disc photons \citep[e.g.][]{mehdipour11a,done12a,digesu14a} have provided viable explanations for type I AGN. \\

A recent breakthrough in our understanding of the nature of the hard excess 
was made with {\it Suzaku}: \citet{tatum13a} revealed that a majority 
of nearby type I AGN show a strong `hard excess'
above 10\,keV, whereby the luminosity observed above 10 keV is too high to be explained by any models previously fit to the data below that energy. 
If interpreted in terms of absorption partially covering 
the line-of-sight, this would require that even the type I AGN are covered 
by low ionization Compton-thick matter, with columns in excess of $N_H>10^{24}$\,cm$^{-2}$ 
covering at least 50\% of the central X-ray emission region. This is in 
contrast to what would be expected in AGN Unification Schemes \citep{antonucci93a}, where the 
sight-lines towards type I AGN are thought to be largely unobscured  \citep[see also][]{burtscher16a} .

However, the observations 
may be explained if the obscuring medium is inherently clumpy \citep{elitzur06a}.   Interest in clumpy reprocessor models has  grown recently on the basis of data in many wavebands. Methods of testing models for the cloud ensemble have been explored using a Bayesian inference framework and applied successfully to Centaurus A  \citep{asensio09a}. Such clumpy models have  also been applied to explain spectral energy distributions \citep[SEDs: ][]{asensio13a,alonso-herrero11a} and  the IR properties of AGN \citep{ramos14a}. 

Popular models for AGN data below 10 keV have suggested the   
X-ray spectral form and variability below 10 keV can be explained by a partial-covering 
absorber (\citealt{pounds04a,pounds04b}, \citealt{terashima09a,mizumoto14a}) and so this hard X-ray model is a natural extension of that picture to higher column densities and energies.  In such models the absorption signature should be calculated along with the scattered (reflected) X-ray emission expected from clumps of the gas that lie out of the line-of-sight \citep[e.g.][]{miller13a}. 
 While the location of the Compton-thick component is not yet known, 
spectral variability timescales show it to be consistent with an origin at or inside a clumpy obscuring torus. Possibilities within the torus include  either an inhomogeneous disk wind or  outflowing clouds that may be part of the 
optical broad-line region \citep[e.g.][]{risaliti07a,ricci10a}. From consideration of the covering fraction and of the kinematics of X-ray absorbing gas, it seems likely that this zone may be part of a clumpy disc wind \citep{proga04a}.   Indeed, models for a Compton-thick wind \citep[e.g.][]{sim10b,proga04a} have shown promise to explain the hard X-ray form and Fe K$\alpha$ emission of local Seyfert galaxies.  Models by \citet{sim10b} and \citet{digesu13a} are examples of some of those that explore the line broadening that can be produced by such outflowing gas.  

If, instead, the hard excess arises through X-ray reflection off Compton-thick matter 
out of the direct line of sight, then its strength measured 
by {\it Suzaku} in many AGN far exceeds what would be expected from a disc  
(or torus) subtending $\sim2\pi$\,sr. This model requires the intrinsic 
X-ray continuum to be heavily suppressed with respect to the reflected emission. This 
could be achieved if the trajectories of the individual continuum 
photons are curved away from the observer towards the plane of the disc  
due to the strong gravitational field of the black hole -- subsequently 
this was referred to as the ``light-bending'' model \citep{miniutti04a}.  Here, a strong, highly relativistically blurred reflection component can be produced which can dominate the intrinsic continuum 
above 10\,keV. Application of this model to AGN spectral data, including 1H 0419-577, \citep{walton13a}, required blurred reflection from material emitted predominantly from 
a few gravitational radii, around a maximal Kerr black 
hole  to explain the hard excess   \citep[e.g.][]{fabian05}. 

The Seyfert 1 galaxy, 1H 0419-577 (z=0.104) provided the first report of the hard excess  phenomenon \citep{turner09b} in {\it Suzaku} data  and turned out to have typical X-ray absorber properties in the context of the later \citet{tatum13a} study.  
1H 0419-577 is also notable for having a highly variable X-ray spectrum in the 1-10 keV band, successfully parameterized in the past using variations in a complex X-ray absorber \citep{pounds04a,digesu14a}.   A correlated variability event in the X-ray and UV bands, studied using {\it XMM} data \citep{pal17a} helps motivate a broad-band study of the variability and spectral form of this AGN. 

{\it BeppoSAX}  provided the first evidence for  1H 0419-577 to have a flat component in the hard 
X-ray regime, \citep[modeled in that case as a $\Gamma \sim 1.5$ powerlaw, ][]{guainazzi98a}; however, it was not until the {\it Suzaku} observations that this was identified as requiring a new Compton-thick absorption component, necessitating a revision of the intrinsic continuum level, and a paradigm-shifting reconsideration of the importance of Compton thick absorption in type 1 AGN \citep{turner09b}.  
 Alternative parameterizations of the hard component as blurred reflection in 1H 0419-577, for example, \citet{walton10a} and \citet{pal13a},  require blurred reflection from a disc extending down to $\sim 1.4 r_g$, with an emissivity index $q > 8$: the high degrees of blurring these models then reduce the disc component  to a smooth continuum form. 

Following on from {\it Suzaku}, {\it NuSTAR}  has provided an opportunity for significant progress in understanding the hard spectrum of AGN, providing data with an unprecedented spatial resolution via imaging optics of 18$''$ FWHM and spectral resolution 400 eV (FWHM) at 10 keV, along with a relatively high effective area up to $\sim 80$ keV. Thus compared to previous non-imaging hard X-ray detectors, {\it NuSTAR} is not background dominated at energies above 10\,keV and is able to provide sensitive measurements of 
the hard X-ray emission from AGN.   Thus {\it NuSTAR} affords an opportunity to disentangle the relative contribution of reflection, absorption as well as Comptonization to provide a self-consistent explanation for the shape of the broad band data.

In this paper we present a $\sim$170 ks {\it NuSTAR} observation of 1H 0419-577.  
This observation  was made with an overlapping exposure from {\it Swift} to  enable a test of physically realistic models for the broadband UV to hard X-ray continuum and reprocessing components in the target source.  
The primary motivation of the observations was to determine the relative importance of Compton thick 
absorption, reflection and Comptonization in shaping the hard X-ray emission from this AGN.
Here, we also consider the archived {\it Suzaku} data to extend the XUV model to account for the marked spectral variability 
observed in 1H 0419-577 between different epochs \citep[e.g.][]{guainazzi98a,page02a,pounds04a}. 
We also use the archived {\it XMM} observation that caught this source in an extreme  low state \citep{pounds04a}, to test the extension of the final model across the full range of behavior exhibited by this AGN. 

\section{Observations}

\subsection{NuSTAR}

{\it NuSTAR} carries  two co-aligned telescopes containing Focal Plane Modules A and B (FPMA, FPMB; \citealt{harrison13a} ) covering a useful bandpass of $\sim 3-80$ keV for AGN. {\it NuSTAR} 
observed 1H 0419-577  on 2015 June 03.  Data were processed with {\sc heaosft 6.16} and the {\it NuSTAR} Data Analysis Software package v. 1.4.1. Event files were created through the {\sc nupipeline} task then calibrated with files from {\sc caldb  20150316} and filtered using standard criteria. \footnote[1]{Upon revision of the paper, we re-extracted the {\it NuSTAR} data with the latest {\sc heasoft} (v6.22) and the {\it NuSTAR}  CALDB dated 20170817 and found that the spectra were consistent within errors with the original spectral products extracted in 2015}.
Source and background spectra were extracted through  regions of  $50''$ radius then binned to  50 counts per spectral channel.  This binning ensured sufficient counts in all channels for fitting using the chi-square statistic. After grouping, 
the data were still sampled more finely than the spectral resolution of the instruments.  In the plots the data are binned more coarsely than allowed in the fit, for visual clarity. 
The net exposure times were about 170 ks for each of FPMA and FPMB. 
FPMA  gave $0.386\pm 0.002$ ct/s from the source while FPMB gave  $0.379\pm0.002$ ct/s, the background level was $< 3\%$ of the total count rate. 
These rates correspond to a 10-50 keV flux  $2.1 \times 10^{-11} {\rm erg\, cm^{-2} s^{-1}}$. Examination of the {\it NuSTAR} images confirmed that no strong hard X-ray source exists that would have compromised the measurement of the hard X-ray spectrum in previous PIN \citep[e.g.][]{turner09a} or {\it BeppoSAX} PDS observations \citep{deluit03} of 1H 0419-577. 

\subsection{Swift}

{\it Swift} observations were obtained  2015 June 03 - 09 (OBSID 00081695001), simultaneous with the {\it NuSTAR} observation. 
The {\it Swift} X- Ray Telescope   \citep[XRT,][]{burrows05a} observations  totaled 4.1 ks and were performed in photon counting mode. The data were reduced using  {\sc heasoft 6.13} with {\sc xrtpipeline} v 0.13.2. Source and background photons were extracted using {\sc xselect} version 2.4c, from circular regions with radii of $47''$ and $189''$, respectively. The XRT yielded a source count rate of $0.48\pm 0.01$ ct/s, and the background rate was $<0.5\%$ of the total. Pileup is negligible in these data. The spectral data were re-binned to have at least 25 photons per bin  (still sufficient for application of the chi-squared statistic, but in this case, application of a higher photon count per bin would have reduced the spectral resolution of the data.)

The {\it Swift} UV-optical Telescope \citep[UVOT][]{roming05a} was used with the U-band filter,  that was the only UV data simultaneous with the 2015 X-ray data. The U filter has a central wavelength 3465 \AA. Source photons were selected from a circle of radius $5''$, yielding a count rate $81.64\pm1.81$ ct/s. The sky background was $< 4\%$ of the total count rate. Using the host galaxy template of \citep{bentz09b}, and following their relation between galaxy luminosity and the  5100 \AA\,  luminosity of the AGN, we scaled the host template to the 5100 \AA\, flux of 1H 0419-577 and from that we estimated the contamination from the host galaxy to be at the $\sim 2\%$ percent level in the U-band, and therefore negligible. 

\subsection{Suzaku}

Four {\it Suzaku} telescopes focus X-rays on to the X-ray Imaging Spectrometer CCD 
array \citep[XIS][]{koyama07, mitsuda07}. XIS units 0,2,3 are front-illuminated (FI) and cover $\sim 0.6-10.0$ keV 
with energy resolution FWHM $\sim 150\,$eV at 6 keV. 
XIS2 developed a charge leak in 2006 November, and has not been used since then.  
XIS 1 is a back-illuminated CCD, giving it an enhanced soft-band response, extending down to 0.2 keV 
but this has a lower effective area at 6 keV than FI CCDs  and has a higher 
background level at high energies.  
{\it Suzaku} also carries a non-imaging, collimated  Hard X-ray Detector  
\citep[HXD][]{takahashi07} whose PIN instrument provides useful AGN data in the  $\sim 15-70$ keV band. 

The {\it Suzaku} observations of 1H 0419-577 were made 2007 July 25 (OBSID 702041010) and 2010 Jan 16 (OBSID 704064010). 
During the 2007 observation the source was observed at the nominal 
centre position for the XIS, during the 2010 observation it was observed at the nominal centre position for the HXD. 
We ran the pipeline processing using  {\sc HEAsoft}  v 6.16 and {\sc caldb 20150316}, following  the standard reduction as previously published for the 2007 observation, 
detailed by \citet{turner09a} . 
% to 
 %exclude data during and within 500 seconds of  entry/exit from the 
%South Atlantic Anomaly, with an Earth elevation angle less than 10$^\circ$ and with 
% cut-off rigidity $>6$ GeV. 
% FI CCDs were in $3 \times 3$ and $5 \times 5$ edit-mode and normal clocking mode and 
%events with grades 0,2,3,4, and 6 were selected, removing hot and flickering pixels using  
%{\sc sisclean}.   Spaced-row charge injection 
%was used. 
The summed exposure time was 411 ks for the FI CCDs during 2007, and 246 ks during 2010.
XIS spectra and light curves were extracted from circular regions of radius 2.9\arcmin \,  with 
background spectra extracted from a region of the same size but offset from the source and avoiding chip corners
where calibration source data are recorded.  
 Data in the range 1.78-1.9 keV were excluded from XIS due to 
uncertainties in calibration around the instrumental Si K edge. 
%Response and ancillary 
%files were created using {\sc xisrmfgen} and {\sc xissimarfgen}. 
The background level was $1\%$ of the total XIS count rate. 

1H 0419-577 is too faint to be detected in the HXD GSO instrument, but was 
detectable in the PIN. The PIN data reduction and a comparison of two methods of PIN background estimation reduction are detailed 
by \citet{turner09a}, and here we followed that method and 
%PIN data were screened to remove data within 500 s of an SAA passage and having day/night elevation 
%angles $> 5^{\rm o}$. We 
used the model ``D'' PIN background 
\footnote[2]{http://www.astro.isas.jaxa.jp/suzaku/doc/suzakumemo/suzakumemo-2007-01.pdf}.  The net PIN exposures were142 ks  and 105 ks during 2007 and 2010, respectively.  1H 0419-577  comprised $\sim 15\%$ of the total source-plus-background counts in the PIN band.  
 For {\it Suzaku}, spectral fits used the only operational FI CCDS, XIS 0 and 3 over 
$0.6-10$\,keV and PIN data in the 
energy range $15-40$\,keV. 
(XIS1 was not used owing to the relatively 
high background level at high energies.) 
%The source data screening time filter was 
 %applied to the PIN background events 
%model file to give a background level appropriate to the parts of the orbit for which we have PIN data. 
%{\sc hxddtcor} was run to apply a deadtime 
%correction to the PIN source spectrum.  The 
%flux of the CXB  is $8 \times 10^{-12} {\rm erg\, cm^{-2}s^{-1}}$ in the 
%15-50 keV band 
%\citep{boldt87a,gruber99a}  and this component was scaled to the PIN field-of-view then combined with the PIN instrument background file to create an total background file. 

During 2007 1H 0419-577 yielded full-band (0.6-10 keV XIS and 15-70 keV PIN, background subtracted) count rates 
1.157 $\pm 0.002$ (per XIS unit)  and $3.053 \pm0.13 \times 10^{-2}$ (PIN) ct/s corresponding to a 
0.5-2 keV flux 
$1.5 \times 10^{-11} {\rm erg\, cm^{-2} s^{-1}}$, 
2-10 keV flux 
$1.8 \times 10^{-11} {\rm erg\, cm^{-2} s^{-1}}$ and 10-50 keV flux  $2.9 \times 10^{-11} {\rm erg\, cm^{-2} s^{-1}}$, 
consistent with the hard flux measured by {\it NuSTAR} when mission and instrument cross calibration 
uncertainty is taken into account.  

During 2010 the rates were 
0.790 $\pm 0.002$ (per XIS)  and $2.71 \pm0.15 \times 10^{-2}$ (PIN) ct/s corresponding to a 
0.5-2 keV flux 
$1.2 \times 10^{-11} {\rm erg\, cm^{-2} s^{-1}}$,  
2-10 keV flux 
$1.4 \times 10^{-11} {\rm erg\, cm^{-2} s^{-1}}$ and 10-50 keV flux  $2.4 \times 10^{-11} {\rm erg\, cm^{-2} s^{-1}}$. 
  
\subsection{XMM-Newton}

{\it XMM-Newton}  (hereafter {\it XMM}) captured 1H 0419-577 in an historical low  state during 
2002 Sep (OBSID 0148000201).  The flux range observed with {\it XMM-Newton}  \citep{pounds04a,page02a} was 
$0.2 - 1.3 \times 10^{-11} {\rm erg\, cm^{-2} s^{-1}}$ in the 0.5 - 2 keV band and 
$0.9 - 1.6 \times 10^{-11} {\rm erg\, cm^{-2} s^{-1}}$ in the 2-10 keV band. 

 The EPIC pn spectrum plus the OM U-band filter data extracted via {\sc SAS v16} were used as a test of the extension of the model to the most extreme flux states.  As the optical and ultraviolet fluxes are variable by up to 30\% for 1H 0419-577 \citep[e.g.][]{pal17a}, we chose only the simultaneous U-band photometry point from the OM data, for the most meaningful comparison. 

Following 
\citet{page02a} the data were screened to limit acceptable event patterns to the range 0-4. Hot and bad pixels were removed. Source counts were obtained from a circular region of $45''$ radius centred on 1H 0419-577, with the background being taken from a similar region on the same chip but offset from the source.  Periods of high background were removed from the observation.  The resultant spectrum had a net exposure of 12 ks.    The pn spectrum was  binned to a minimum of 50 counts per energy bin and the binned data maintained sampling finer than the spectral resolution of the pn.

\section{Spectral Fitting Results}

The full-band {\it Suzaku} XIS light curve showed  $< 40\%$ flux variability during 2007 and 2010. 
The source flux variations were $<20\%$ during the {\it NuSTAR} observation. The source has previously been reported as 
showing a steady light curve during the {\it XMM} observation of 2002  Sep \citep{pounds04b}. 
Thus, in this paper we report only the mean spectra from each epoch observed. 

Spectra are analyzed with XSPEC v12.9.0. All errors correspond to 90\% confidence levels for one parameter of interest.

\subsection{Cross calibration and grouping}

When performing joint {\it NuSTAR} fits with the {\it Swift}  XRT and UVOT data, we used {\it NuSTAR} data over 3-70 keV, and {\it Swift} XRT over  0.3-6.0 keV. We introduced a multiplicative factor to account for differences in the relative flux calibrations of the various instruments. Fixing a constant model parameter to 1 for FPMA, we allowed the normalization of FPMB and the {\it Swift} XRT and U-band spectral files to vary  relative to that. For the 2015 data we found the cross-normalization factor was $< 5\%$ between the 2015 datasets, and so we then constrained the relative normalizations to lie within that range, because this later sped up the more complex multi-epoch fitting.

In the spectral model,  the PIN model flux was increased by a 
factor 1.16 for 2007 and 1.17 for 2010, as recommended, to account for the instrument  
cross-calibration at the epoch of those observations   (and these cross calibration factors were fixed in the fit) 
 \footnote[3]{ftp://legacy.gsfc.nasa.gov/suzaku/doc/xrt/suzakumemo-2008-06.pdf}. 
%A  2\% systematic error was applied to the PIN background spectrum before spectral fitting. 
XIS data were binned at the HWHM resolution for each instrument while PIN data were binned to be a minimum of 5 $\sigma$ 
above the background level for the spectral fitting.  Note that as the PIN data are of much lower signal to noise compared to {\it NuSTAR}, in the subsequent plots we have binned the PIN data to give a single 
flux point compared to the {\it NuSTAR} spectra.

 For  {\it XMM} the EPIC pn spectrum plus the OM U-band filter data extracted via {\sc SAS v16} were used as a test of the extension of the model to the most extreme flux states. 
A 10\% error was applied to the OM U flux, to account for the flux calibration uncertainty of the OM  \footnote[4]{http://xmm2.esac.esa.int/docs/documents/CAL-TN-0019.pdf}. 

\begin{figure}
%\epsscale{.10}
%\includegraphics[width=11.5cm, height=6cm, angle=0]{fig1.pdf}
\includegraphics[scale=0.43,width=8cm, height=8cm,angle=-90]{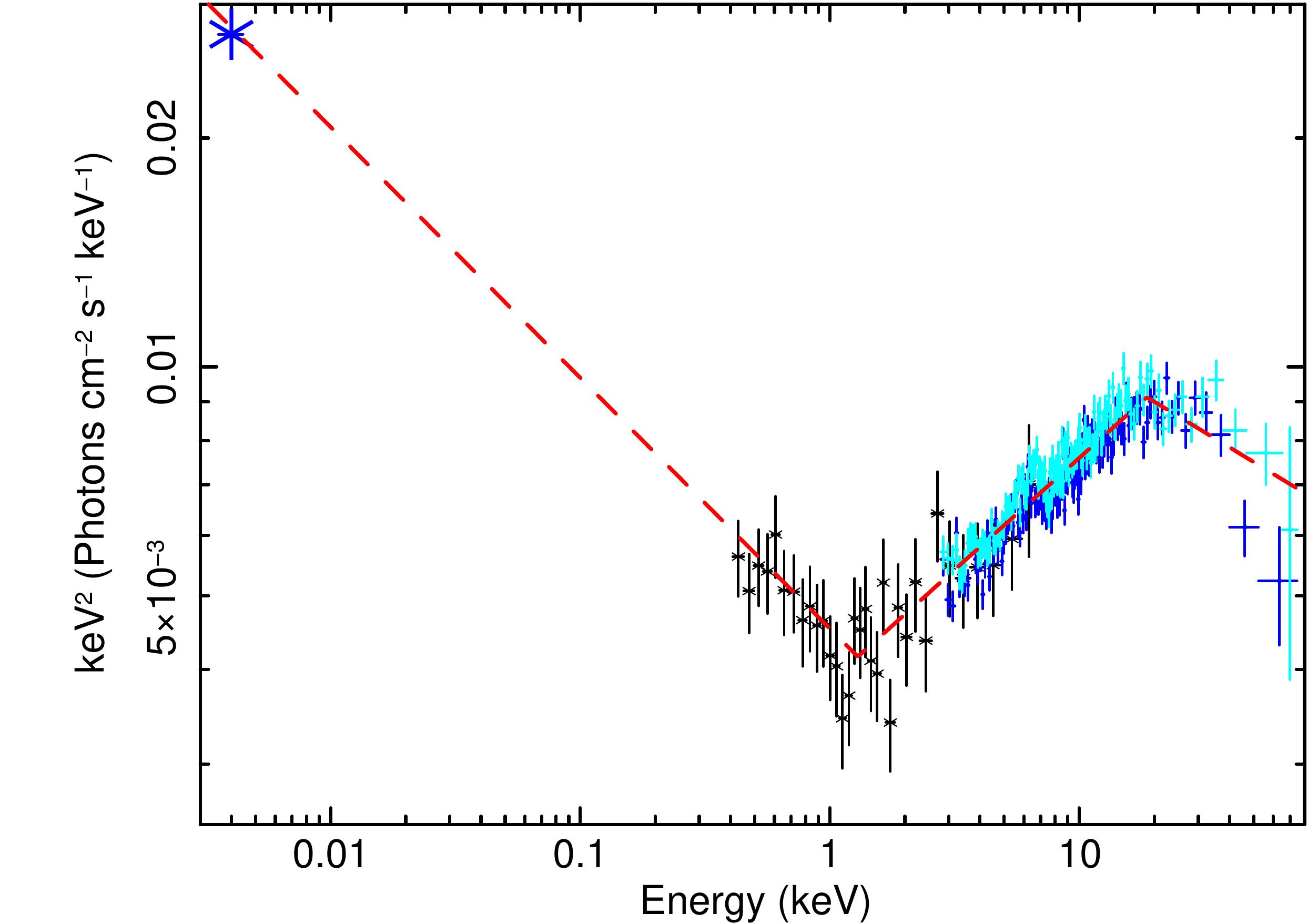}
\caption{The simultaneous {\it NuSTAR} and {\it Swift} SED of 1H\,0419-577. A double broken power-law model (red dashed line) has been fitted to the 2015 data, showing the {\it Swift} UVOT (U band, dereddened, dark blue star) and XRT (black cross) data, along with the {\it NuSTAR} FPMA (dark blue cross) and FPMB (aqua cross). Note the hard X-ray continuum ($\Gamma=1.7$) up until a break energy of 20\,keV, above which the 
spectrum significantly steepens ($\Gamma=2.2$). A strong UV to soft X-ray excess is present below 1 keV. } 
\label{fig:bknpow}
\end{figure}

\subsection{Basic parameterization of the broadband spectral form}

A full covering screen of neutral gas was included in all fits presented in this paper, to represent the Galactic line-of-sight column density 
${\rm N_H} = 1.26 \times 10^{20} {\rm atoms\, cm^{-2}}$, parameterized using {\sc tbabs}. We also accounted for the UV extinction of 
E(B-V)$= 0.0186$ from \citet{cardelli89a} using the {\sc redden} model.  
In {\sc xspec} the transmission is set to unity short-ward of the Lyman limit, allowing {\sc redden} to be used in combination with {\sc tbabs} in the fit.

The flux observed for 1H 0419-577 during the 2015 observation corresponds to an observed luminosity  
$L_{2-10} = 4.1 \times 10^{44}\, {\rm erg\,s^{-1}}$ 
(assuming $H_0=70\, {\rm km\, s^{-1}\, Mpc^{-1}}$).  

First, we attempted to make a basic characterization of the broadband XUV spectral form during the 2015 campaign, to extract a simple phenomenological parameterization of the spectral energy distribution. To this end, we applied a simple double broken power-law model to the 2015 {\it NuSTAR} plus {\it Swift} data over 0.3-79 keV. This provided a good fit to the data with a $\Gamma \sim 2.4$ slope joining the U-band point to the soft X-ray data (Figure ~\ref{fig:bknpow}). At  1 keV this flattens to a $\Gamma \sim 1.7$  slope, continuing up to $\sim 20$ keV. In the 20-50 keV range the spectrum steepens again, to $\Gamma \sim 2.2$, appearing to recover the slope of the soft X-ray to UV portion of the spectrum.   For this epoch the observed $\alpha_{ox} \sim 1.76$.  
%Integrating the flux we find $F_{1-1000 Ryd} \sim 3 \times 10^{45} {\rm erg\, s^{-1}}$.  
%This SED  was used as an input to {\sc xstar}, to generate an emission table used in our fits. 

\begin{figure}
%\epsscale{.60}
%\includegraphics[width=6cm, height=7cm, angle=-90]{fig2.pdf}
\includegraphics[scale=0.33, width=8cm, height=8cm,angle=-90]{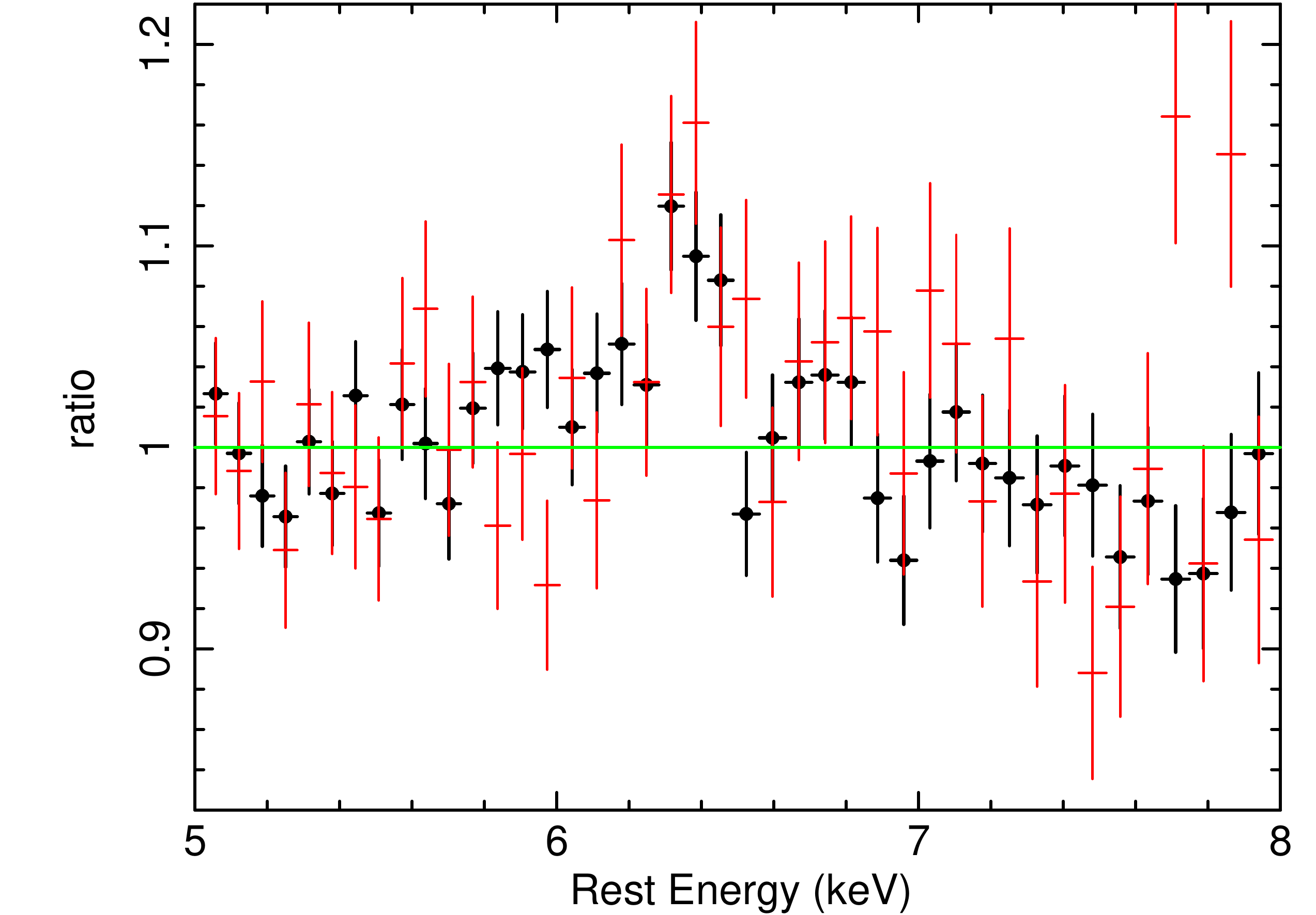}
\caption{Suzaku XIS\,0+3 data in the Fe K regime: 2010 (red) and 2007 (black) compared to a local powerlaw continuum model over 5-8 keV. A weak, but narrow, iron K$\alpha$ emission line is seen 
near to 6.4\,keV in the QSO rest frame.}
\label{fig:fek}
\end{figure}

We also attempted a simple parameterization of the Fe K emission line, known to be weak \citep{turner09a}, to guide our subsequent model selection. 
The {\it Suzaku} XIS data yield the best constraints on Fe K emission (Figure~2). Measuring the line equivalent width relative to a `local' powerlaw continuum (fitted over 3-10 keV), we fit the line with a Gaussian model, obtaining a reduction in fit statistic $\Delta \chi^2 \sim 54$ for 3 additional degrees of freedom (giving $\chi^2=223.7/208\, d.o.f$). The fitted line energy was $E_{Fe}=6.34\pm0.06$ keV, width $\sigma < 380$ eV, and equivalent width $31^{+34}_{-10}$ eV (Figure ~\ref{fig:fek}). 
%The  upper limit on the line  width and shape of the profile indicates that there is no significant relativistic blurring of the line  
Applying the same model to the {\it NuSTAR} data yields a line equivalent width $55\pm15$ eV for a width fixed at $\sigma=100$ eV.

\subsection{Comptonization Models}

The simple double broken power-law parameterization of the data has characterized the basic spectral shape of the source, and suggests that a broadband continuum model should be applied, that allows for  spectral steepening  above 20 keV, similar to those observed previously for AGN in {\it BeppoSAX} data 
\citep[e.g.][]{dadina08a}. 
To investigate the high-energy spectral roll over, we modeled the spectrum using Comptonization models \citep[e.g.][]{petrucci01a}.

 As a previous {\it XMM} RGS exposure of 1H 0419-577  has detected soft-band line emission from  C, N, O and Ne  \citep[e.g][]{pounds04a,digesu13a}, we need to account for that gas before trying to constrain the continuum form. 
 Detailed analysis of the RGS data led  \citet[][ using the SPEX code]{digesu13a} to suggest that the X-ray emitter resides on pc size scales. 
To account for the line emission from that gas we included in the model  and all subsequent models an {\sc xstar} table at
the systemic redshift of the host galaxy. The emission
line table was generated using {\sc xstar} v 2.2.1 with a
turbulent velocity 300 km/s and an illuminating
spectrum consistent with the SED characterization
(Section 3.2). The {\sc xstar} table assumed a density
$n = 10^{12}{\rm cm}^{-3}$, and we
note that the emergent spectrum calculated by 
{\sc xstar} is not sensitive to the density at the spectral
resolution available from {\it NuSTAR} and the {\it Swift} XRT.  

An initial fit of the soft-band XRT spectrum shows that our data do not allow new insight into the soft-band X-ray emitter, but are consistent with the previous results from RGS. As the column density and normalization of this component are degenerate, we fixed the column at 
$10^{21} {\rm atoms\, cm^{-2}}$ for the fit \citep[i.e. in the middle of the range fitted by][]{digesu13a} and allowed the normalization to be free.  An initial fit to the 0.3-2.0 keV band alone yielded an ionization parameter log\, $\xi=1.30\pm 0.51$ (where  $\xi=\frac{L}{nr^2}$ and $L$ is the ionizing luminosity integrated from 1-1000 Ryd, $n$ is the proton density and $r$ the distance of
the material from the central black hole).  This value of $\xi$ was fixed for the soft emitter, in subsequent fits.  
We also note that some of the narrow soft-band emission lines could arise from a distant scattering/reflection component. 

The Fe K$\alpha$ emission line observed is stronger than that which could be produced by the soft-line emitting gas, and so a Gaussian line component was initially added to the fit to account for that. Note that in subsequent fits the Fe K$\alpha$ emission is self consistently fitted using ionized 
reflection models.

With these components in place, we constructed two variations on a simple Comptonization model. Both versions used {\sc comptt} to describe the soft X-ray band, this is an analytic model describing Comptonization of soft photons in a hot plasma \citep{titarchuk94}.  
Then the hard part of the spectrum was tested against two alternative models,   a second {\sc comptt} component or 
a thermally Comptonized continuum, available in {\sc xspec} as {\sc nthcomp} \citep{zdziarski96a,zycki99a}.  
The free parameters of {\sc nthcomp} are the temperature of the corona, the soft-photon temperature and the  power-law index of the spectrum. 
The advantage of {\sc nthcomp} over {\sc comptt} is that the seed photons in {\sc comptt} are limited to have a  black-body distribution, those in 
{\sc nthcomp} can be chosen to have a multi-colour disc distribution. 
The {\it Swift} U band point was omitted from these fits, as neither the {\sc comptt} nor {\sc nthcomp} models extend into the UV region as implemented in {\sc xspec}.  

We assumed a plasma temperature of 2 keV to account for the soft part of the spectrum, and obtained a fitted optical depth 
$\tau=1.9^{+0.4}_{-0.5}$ for that cool zone. For the hard X-ray band, 
parameterization as {\sc comptt}, using an input black body temperature of 10 eV (reasonable for a black hole of this mass, accreting at high efficiency) yields  a plasma temperature  $kT = 15.1\pm0.8$ keV, $\tau = 2.5\pm0.1$, ($\chi^2=957/949\, d.o.f.$), while {\sc nthcomp} yields $kT=13.4\pm1.0$ keV and a photon index $\Gamma = 1.74\pm0.01$ , ($\chi^2=946/949\, d.o.f.$).  

While these are good fits from a statistical perspective, the fitted temperatures are extremely low, yet 
conversely the photon index is unusually hard and thus the overall spectral form may not be viable for an AGN. There is also a concern that the low fitted temperature might be an artifact of the spectral curvature from the known absorber in this AGN, which can modify the continuum so that the observed photon index is much 
flatter below 10\,keV compared to the steeper ($\Gamma>2$) primary continuum which emerges at higher energies. 
Indeed detailed modeling of the spectral form and marked spectral variability sampled over a large body of X-ray data  for 1H 0419-577  provide compelling evidence for the presence of a variable, partial-covering absorber \citep[e.g.][]{pounds04a, pounds04b,page02a} comprising at least two zones of ionized gas.  

% Detailed modeling of the spectral form and marked spectral variability sampled over a large body of X-ray data  for 1H 0419-577  provide compelling evidence for the presence of a variable, partial-covering absorber \citep[e.g.]{pounds04a, pounds04b,page02a} comprising at least two zones of ionized gas.  We therefore added two absorption zones to the Comptonization model above. This improved the fit to 
%$\chi^2=938/947\, d.o.f.$ and yielded a column density $N_H=1.62 \times 10^{22} {\rm cm^{-2}}$,  log $\xi=-0.57^{+0.02}_{-0.09}$ erg\, cm\, s$^{-1}$ covering 38\% of the source, plus $N_H=4.40 \times 10^{23} {\rm cm^{-2}}$ of cool gas, covering 47\% of the source. While this fit provides an acceptable model for the local X-ray continuum and the shape of the high energy turnover, to properly understand this source we need to reconsider the absorption model after first establishing a broader model that can fit the underlying UV to X-ray continuum. 
%We therefore return to these absorption fits in Section 3.5.     

\subsection{Reflection Modeling}

In order to test the applicability of the popular reflection model, we then tested if the higher energy spectral curvature in the {\it NuSTAR} data as well as the soft continuum seen in {\it Swift} XRT could be explained by an ionised reflector. To this end we first fit the spectra using the most recent version of the \textsc{relxill} blurred disk reflection model (Garcia \& Dauser 2014), which convolves the \textsc{xillver} reflection model (Garcia et al. 2013) with the relativistic \textsc{relline} code (Dauser et al. 2013). We assumed Solar abundances and linked the illuminating photon index to that of the primary power-law. We fit the model over the $3-79$\,keV range of the NuSTAR data and the  $\sim0.4-5$\, keV band of the {\it Swift} XRT.

Among the several flavours of the model, we first tested the standard version of \textsc{relxill}, which allows the user to vary the high energy cut-off, which can then be linked to the cut-off energy of the primary power law. The photo ionised emitter, previously required to fit the narrow line emission in the soft X-ray band, is also retained in the model. We fixed the black hole spin to $a=0.998$ (which otherwise is not constrained) and found that the model accounts well for the overall spectral shape (see Figure 3, upper panel), with an overall fit statistic of $\chi_{\nu}^2=940.1/949$. The reflection parameters yield an ionisation state of $\log\xi=2.70_{-0.30}^{+0.04}$\,erg\,cm\,s$^{-1}$, a hard photon index of $\Gamma=1.64\pm0.04$ and a disk inclination of $\theta=26_{-8}^{+9}$.  We note that the model is not dominated by the reflection component and the flux ratio between the reflector and the direct power law is relatively low, with $R=0.21_{-0.10}^{+0.11}$ as measured over the 10-50\,keV band. The reflector does not require a high degree of blurring, with an emissivity index of $q=1.98_{-1.56}^{+1.23}$.  However, the fit still requires the presence of relatively low cut-off energy, $E_{\rm cut}=63_{-9}^{+8} $\,keV. If we fix the cut-off energy at $ E_{\rm cut}=300$\,keV,  the fit is worse by $\Delta \chi^2=60.4$.  We also checked alternative reflection scenarios where the higher energy rollover is a signature of a highly blurred and strong  reflector, but generally these models require a low iron abundance  ($<0.55\times$ Solar) and a rather high emissivity ($q>6$), in order to explain the smooth high energy continuum curvature and the lack of a strong Fe K$\alpha$ emission line. However such blurred reflection models can account for the soft X-ray excess that is present and a separate soft X-ray continuum component modeled with \textsc{comptt} is no longer required in this particular case.  Furthermore, these fits are significantly worse and can be rejected compared to the case above.

\begin{figure}
%\epsscale{.90}
\includegraphics[scale=0.9,width=7cm, height=8cm, angle=-90]{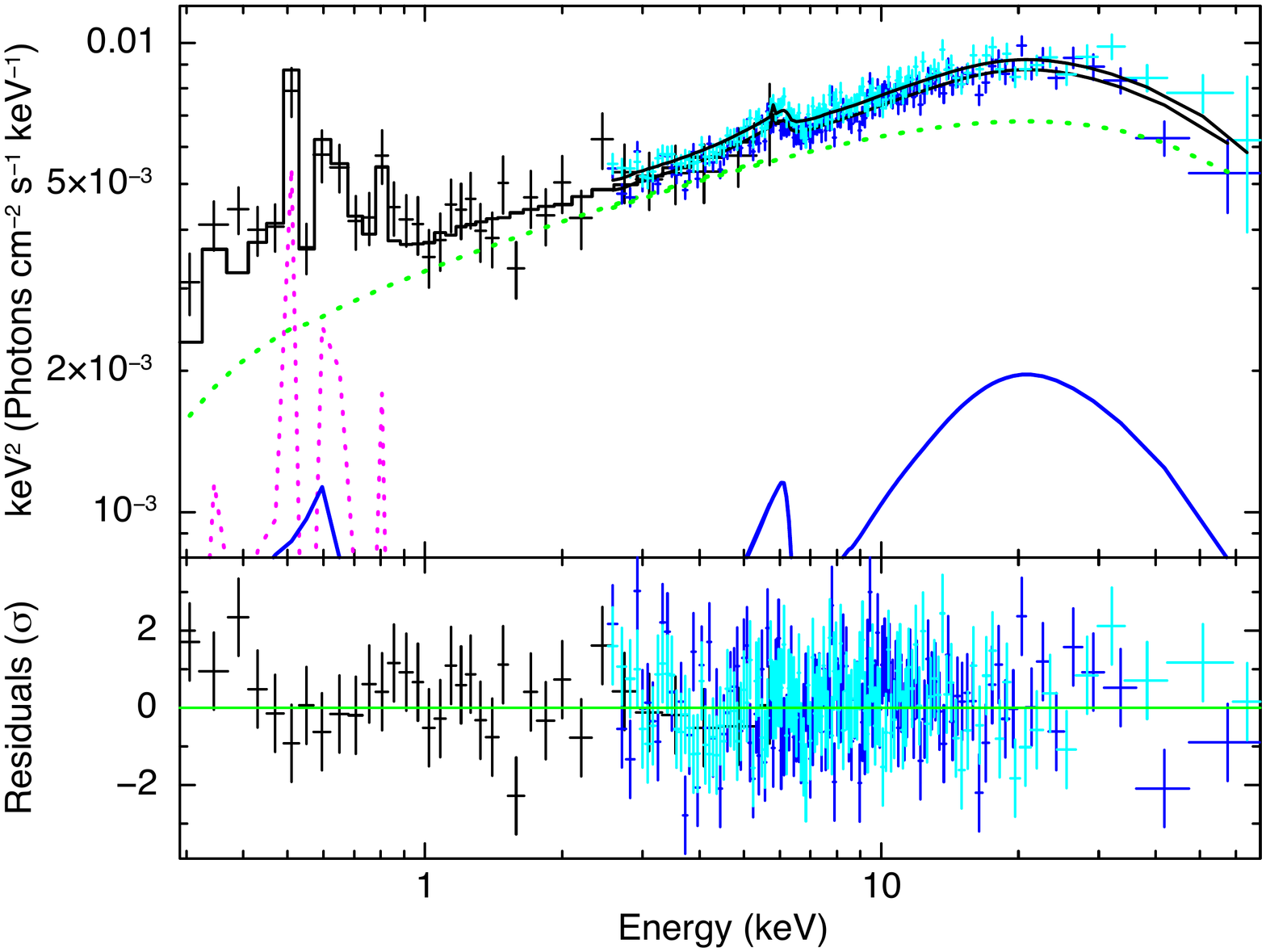}
\includegraphics[scale=0.9,width=7cm, height=8cm, angle=-90]{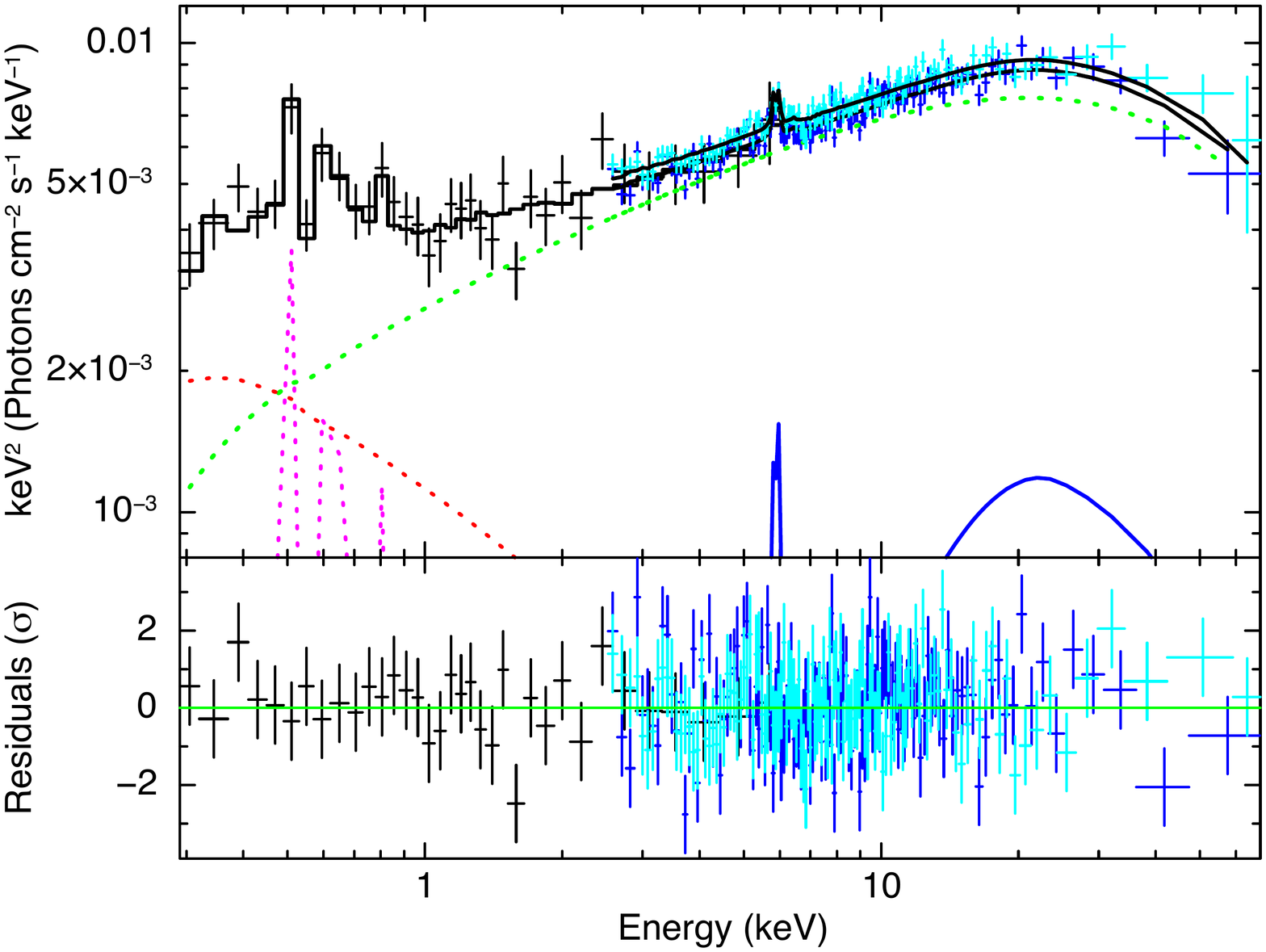}
\caption{Data, model and residuals to joint fits to the {\it NuSTAR} (FPMA dark blue, FPMB aqua)/{\it Swift}(black cross)  data. Top: (a) The model comprises  a power-law continuum (green dashed line) and photo-ionised emitter (magenta dashed line) with blurred reflection modeled using  {\sc relxill} (dark blue solid line) with the cut-off energy allowed to be free.  Bottom: (b) the reflection model, without relativistic blurring (i.e  modeled using {\sc xillver} without convolution with {\sc relline}). Model components as for  a), with the addition of a  {\sc comptt}, shown as a red dashed line. See Section 3.4 for details. }
\label{fig:reflection} 
\end{figure}

\begin{figure}
%\epsscale{.60}
\includegraphics[scale=0.32,width=8cm, height=8cm,angle=-90]{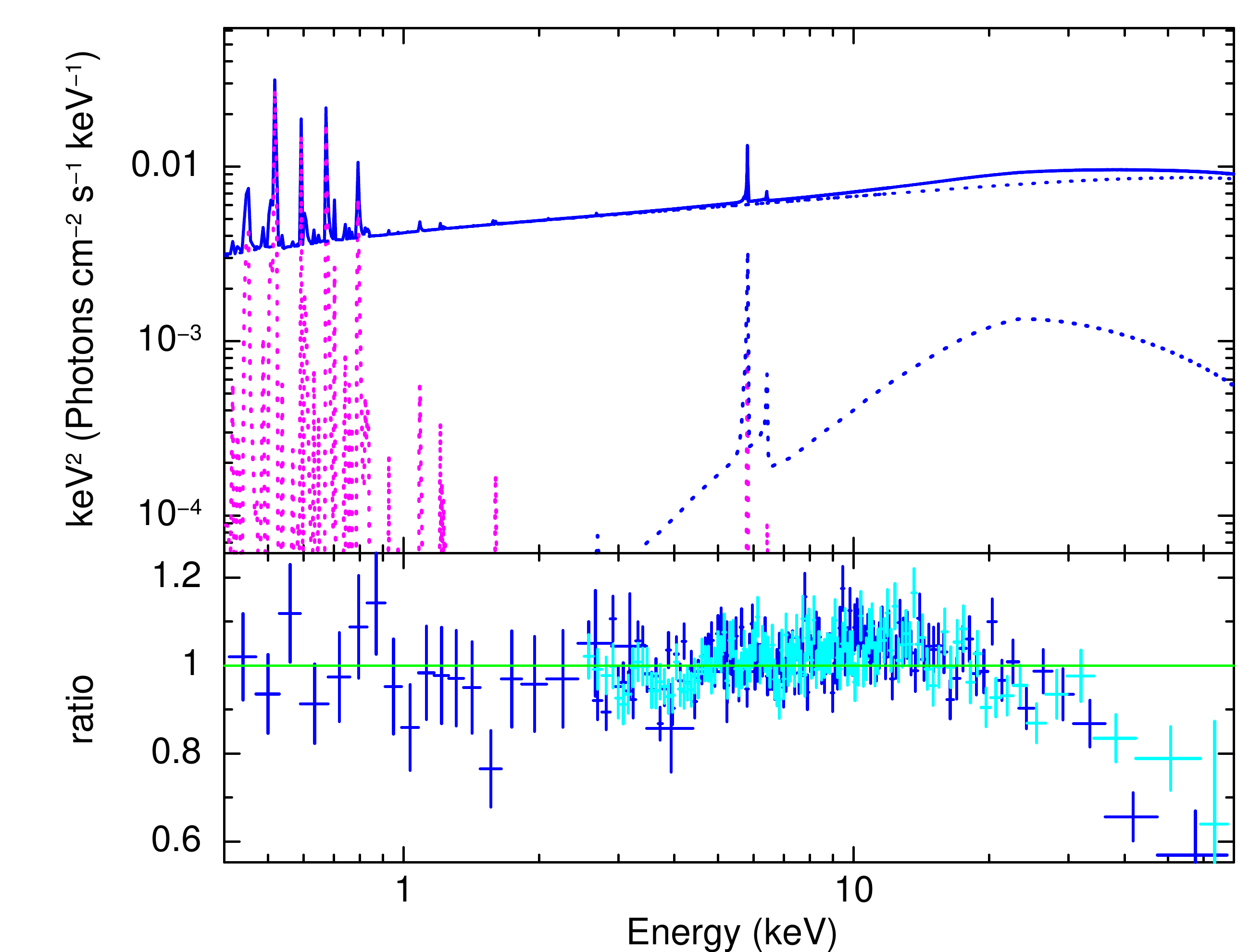}
\caption{The initial fit to {\sc optxagnf} plus {\sc xillver} to the {\it NuSTAR} and {\it Swift} data as described in Section 3.5. The model does not include intrinsic absorption, but does include {\sc optxagnf} (upper dashed blue line) a contribution from distant 
reflection quantified using {\sc xillver}  (lower dashed blue line) and soft band emission quantified using {\sc xstar} (magenta dashed line). The U-band point is fitted in the model, but the 
X-ray data only are shown for clarity. 
The lower panel shows X-ray ratio residuals from FMPA  and the XRT (dark blue) and from FPMB (aqua). Note the strong curvature present in the residuals to the {\it NuSTAR} data.  \label{fig:underpred}}
\end{figure}

Given the low emissivity index for the above reflection fit and the relatively narrow width of the iron K$\alpha$ line (e.g. as implied by the previous Gaussian fits), we also tested whether the spectrum can be simply modelled using an unblurred reflector originating from scattering off more distant material. To test this, we replaced the \textsc{relxill} model with a \textsc{xillver} ionised reflection table (Garcia et al. 2013), with a variable cut-off energy, which simulates the angle averaged reflection spectrum of a power-law illuminating a plane-parallel and optically thick slab. We assumed Solar abundances and again linked the illuminating photon index and high energy cut-off of the reflector to that of the primary hard X-ray power law. The fit statistic is again very good, with $\chi_{\nu}^{2}=934.2/950$, statistically equivalent to the blurred reflection fit and can account for the overall shape of the broad-band spectrum (see Figure 3, lower panel). Again the primary power law has a rather hard photon index ($\Gamma=1.52\pm0.12$) and requires a low cut-off energy of $E_{\rm cut}=44^{+11}_{-12}$\,keV to model the rollover at high energies. The reflector has an ionisation state of $\log\xi=2.3\pm0.4$, while its contribution is also weak compared to the power-law; measured over the $10-50$\,keV band the flux ratio is $R\sim0.12$. Note that unlike for the blurred reflection model, an additional soft continuum component modelled using the \textsc{comptt} model,  is required to account for the soft excess in the {\it Swift} data, as noted in previous analysis (Pal \& Dewangan 2013). In this case, the electron temperature of this soft Comptonisation component has been fixed at $kT=1$\,keV and has an optical depth of $\tau=4.5\pm1.2$.

\begin{figure}
%\epsscale{.90}
\includegraphics[scale=0.9,width=4.85cm, height=8cm, angle=-90]{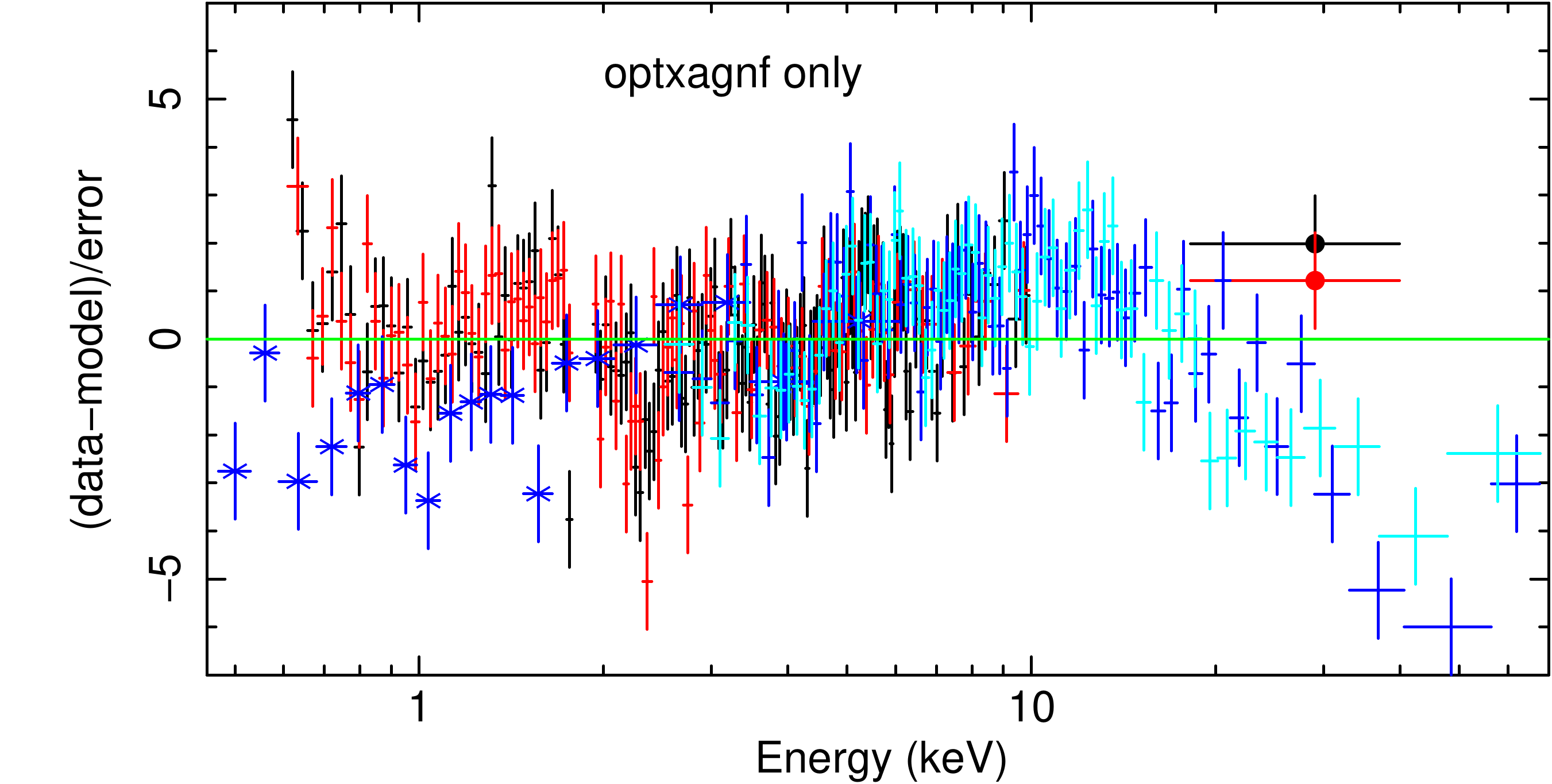}
\includegraphics[scale=0.9,width=4.85cm, height=8cm, angle=-90]{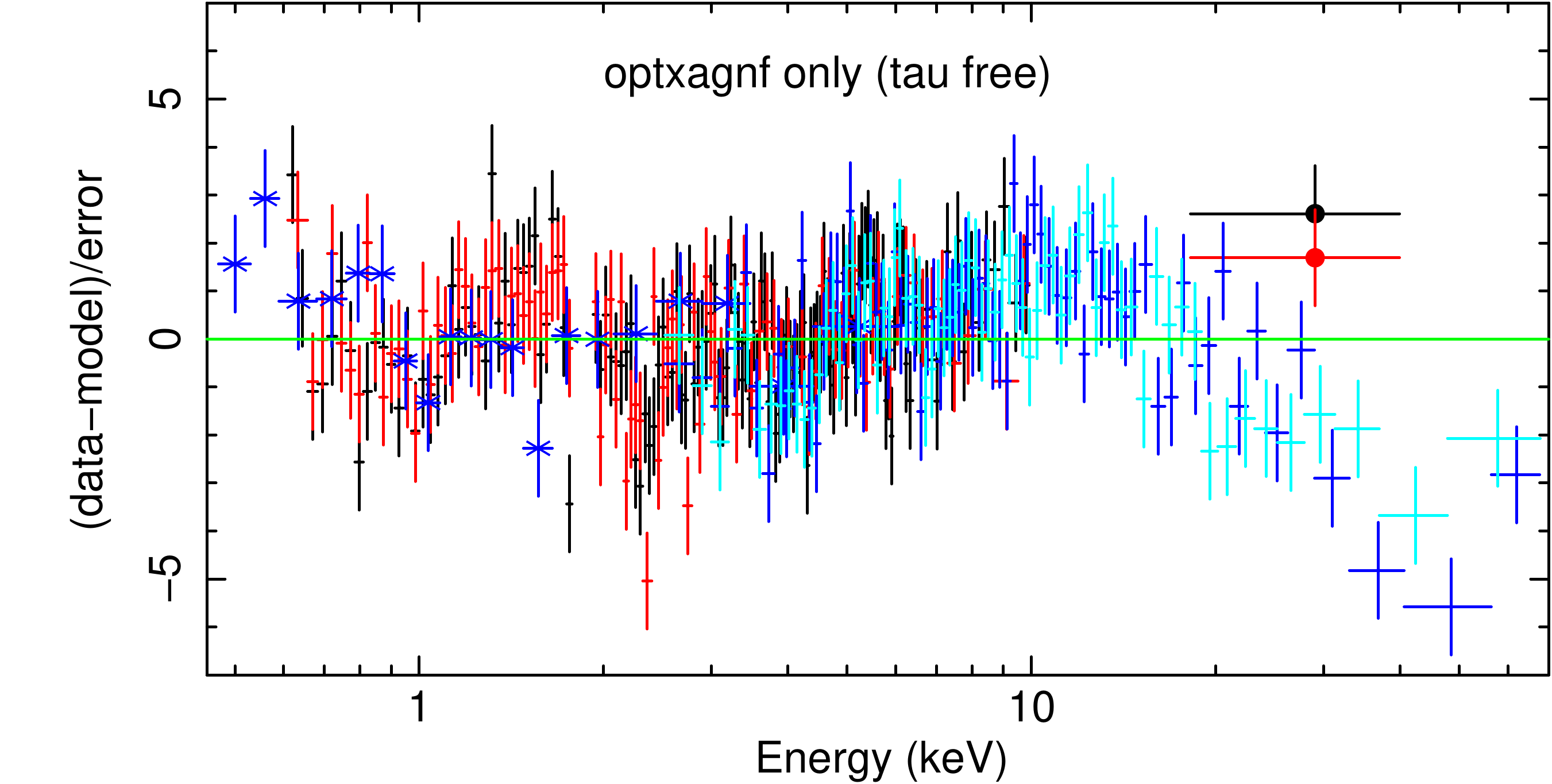}
\includegraphics[scale=0.9,width=4.85cm, height=8cm, angle=-90]{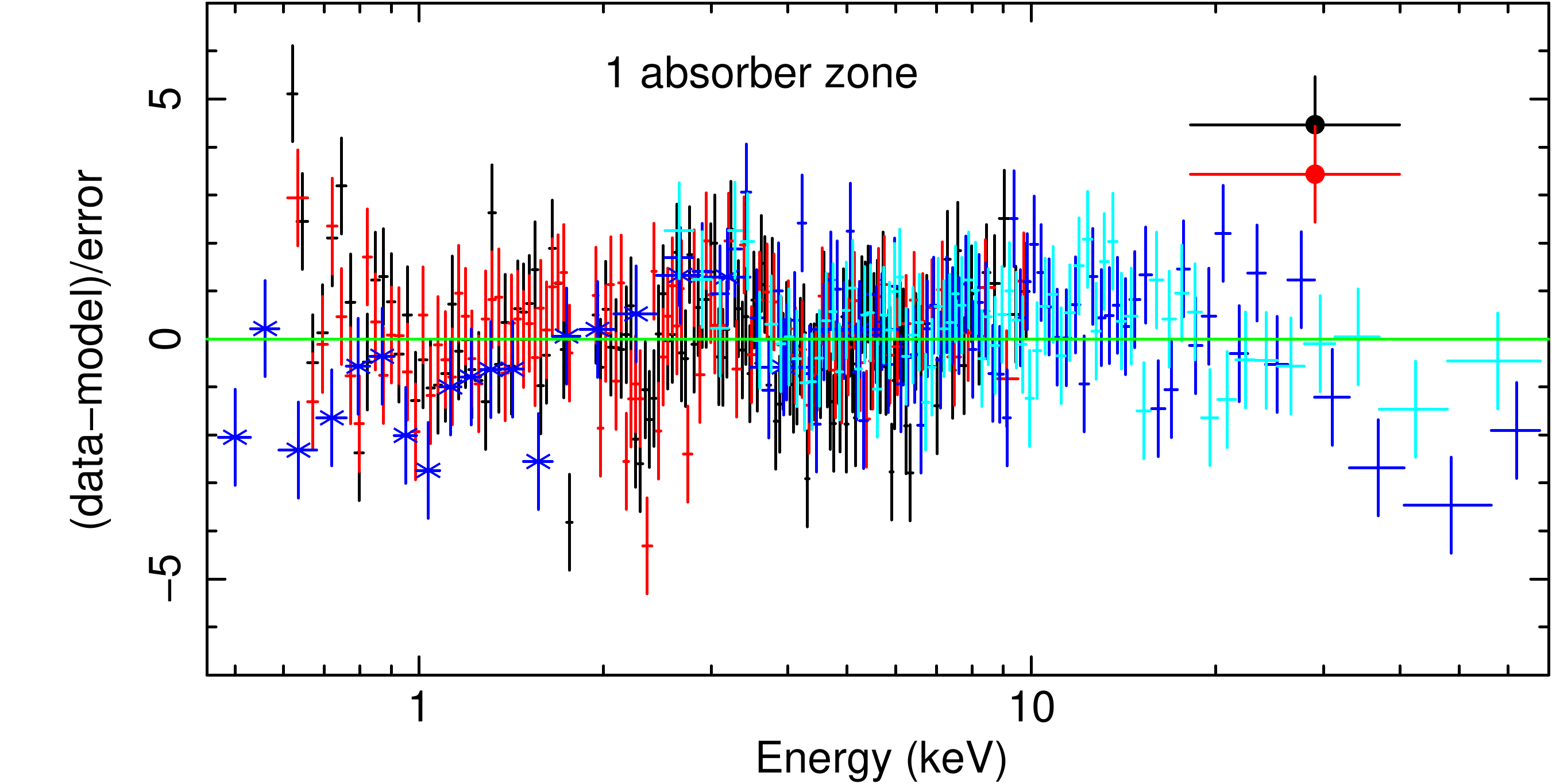}
\includegraphics[scale=0.9,width=4.85cm, height=8cm, angle=-90]{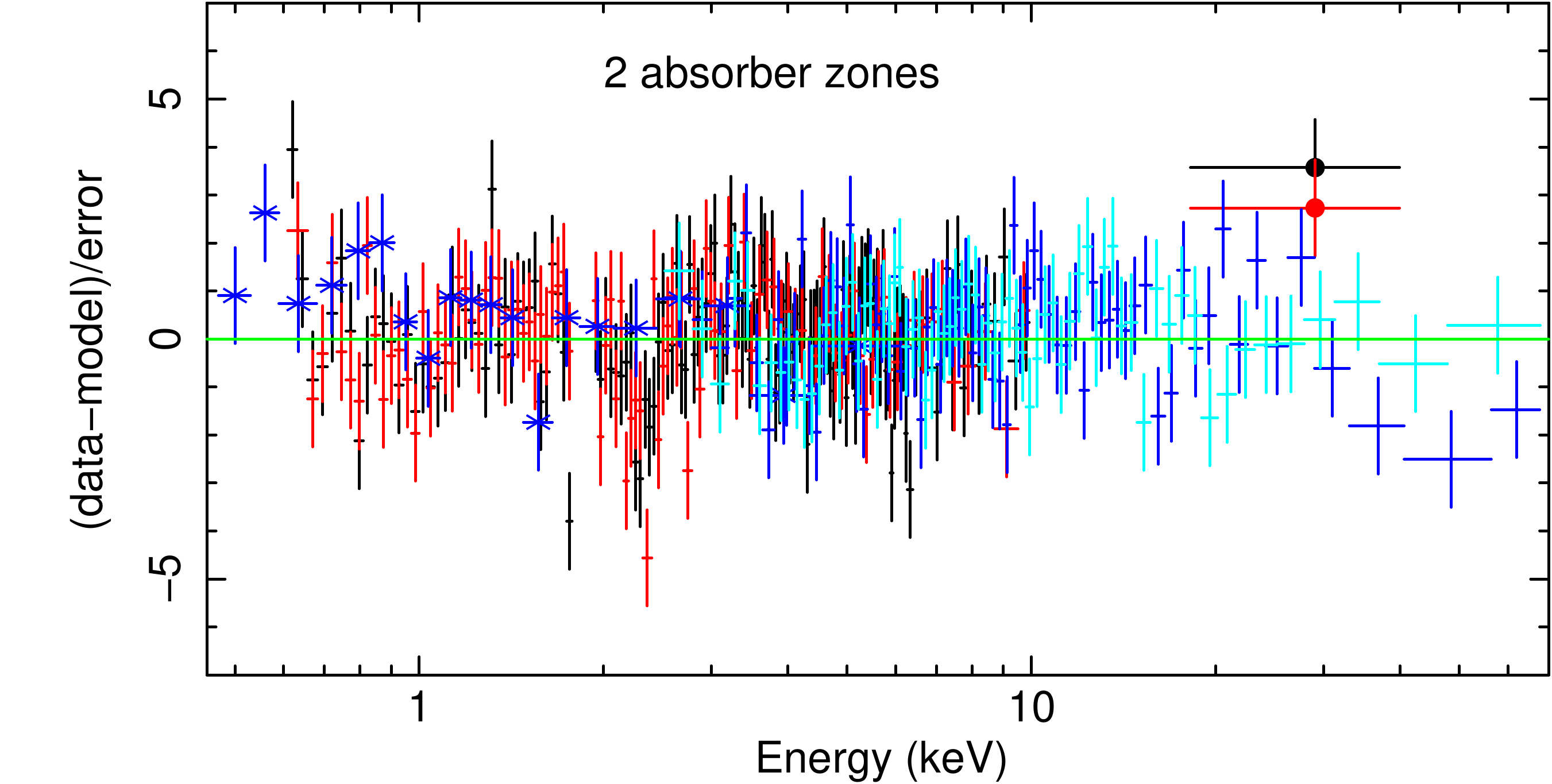}
\caption{Residuals: joint fits to  {\it NuSTAR} (FPMA dark blue, FPMB aqua)/{\it Swift}(dark blue cross) and {\it Suzaku} data (2007 black, 2010 red) using the {\sc optaxagnf} model. Parameters were free but linked  between epochs, see Section 3.6 and Table 1 for details. First only a colourless continuum variability (via the fraction $\frac{L}{L_{Edd}}$) was allowed between epochs (top) with no intrinsic absorption included. Then the optical depth of the warm corona ($\tau$) was allowed to vary between epochs (panel 2). Significant residuals are still present in the spectra that are unaccounted for in the bare coronal continuum model. 
Following this a single zone of absorption was allowed in the fit (panel 3), with covering fraction allowed to vary between epochs. Finally, two zones of absorption were allowed (bottom), accounting for the remaining residuals.  
Note: due to the  lower S/N and higher systematics of the non-imaging {\it Suzaku} PIN data, 
these have been binned to a single 15-40\,keV flux point.}
\label{fig:curvature} 
\end{figure}

We conclude that while reflection models give a good account of the broad band X-ray spectrum of 1H\,-410-577, they cannot explain the hard X-ray curvature without a value for the high energy cut-off that lies within the {\it NuSTAR} bandpass. We find that the fitted cut-off  is lower than typical values in many AGN (Fabian et al. 2015). Furthermore either a moderately blurred or a distant reflector can equally provide an acceptable fit of the reflection features in the data and thus highly blurred reflection is not required to model the broad band spectrum. To understand this source we need to consider physical models that can fit the underlying UV to X-ray continuum, as well as to investigate the effect of any absorption on the primary continuum emission, which can impart significant spectral curvature upon the primary continuum. We now consider this in the next Section.

\subsection{The {\sc optxagnf} model}

To account for the broadband UV to X-ray continuum form using an accretion disc framework, we then turned to the {\sc optxagnf} model \citep{done12a}. In this model the gravitational energy released in the accretion disc  is emitted as a black-body with colour-temperature correction, whose temperature depends on radius.  This black body extends to 
the so-called  `coronal radius',  ${\rm R_{cor}}$ (expressed in gravitational radii, $r_g=\frac{GM}{c^2}$ where M is the mass of the black hole in solar masses, G is the gravitational constant and c is the speed of light). Below ${\rm R_{cor}}$ the system energy cannot fully thermalize and is distributed between low energy and high energy electron populations, giving rise to soft and hard Comptonization components, forming the so-called soft excess and hard-band powerlaw, respectively. 
The hard Comptonization component is parameterized by a power-law with cutoff energy of 100 keV.  \citet{done12a} normalise the model by $\frac{cos i}{cos 60^o}$ as the optical emission is fairly isotropic for inclination angles $i < 60^o$. 
The {\sc optxagnf} model is compelling because it allows us to make a physically meaningful connection between the UV and X-ray emission, sampled by the simultaneous {\it NuSTAR} and {\it Swift} data.  As the Galactic line-of-sight column is low in the direction of 1H 0419-577, this object is well-suited for fitting with a broad SED model such as {\sc optxagnf}.

\begin{figure}
%\epsscale{.60}
\includegraphics[scale=0.32, width=7cm, height=8cm, angle=-90]{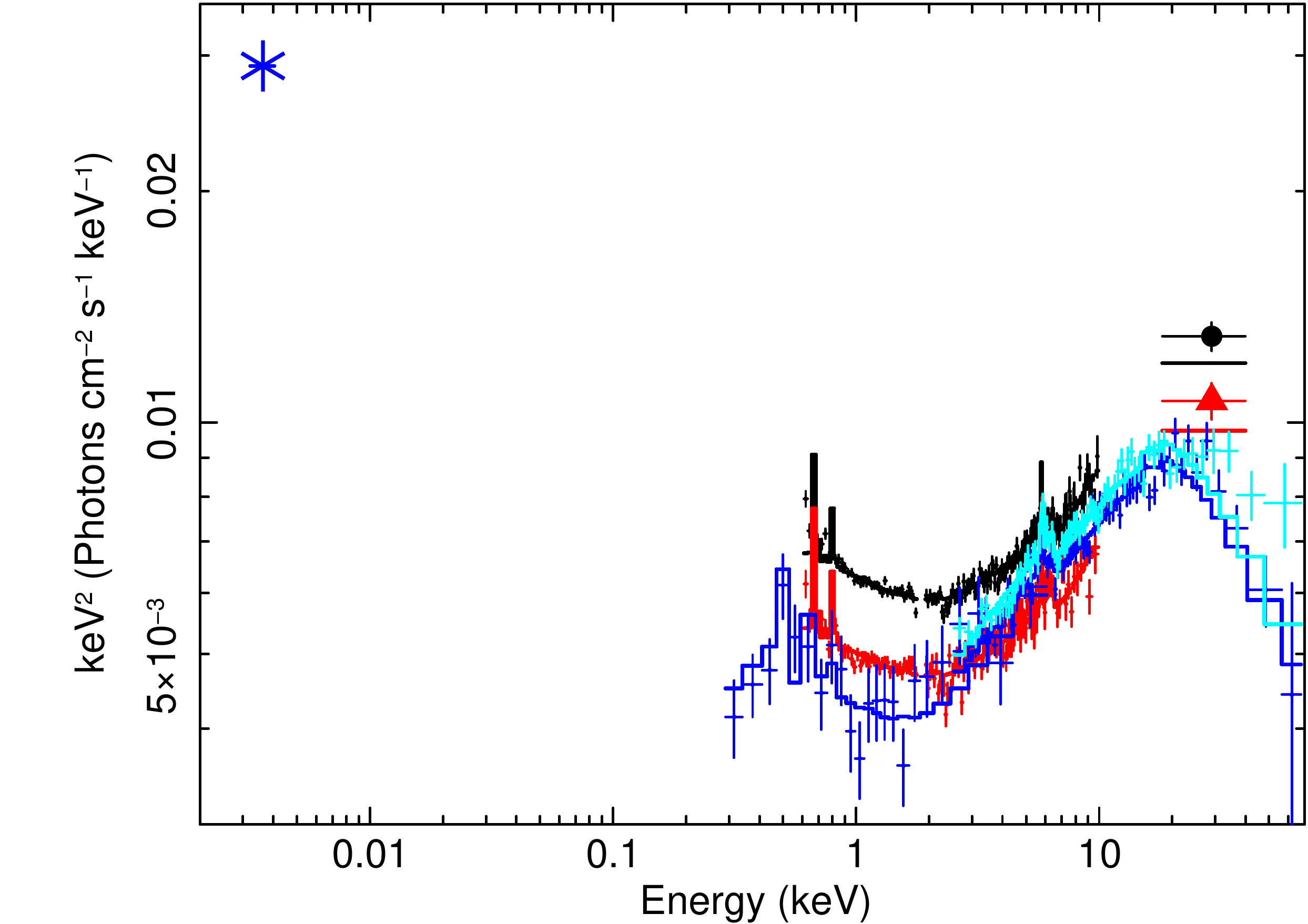}
\caption{Unfolded data versus the best-fit multi-epoch {\sc optxagnf} model from Table~1:  Symbols as for Figure~5. 
%2007  (black, {\it Suzaku}),  2010  (red, {\it Suzaku}), 2015 (blue, {\it Nustar, Swift}). 
Note that for the {\it Suzaku} PIN data, the model is corrected for the calibration cross normalization factor (relative to XIS), but the data are not corrected. \label{fig:nustar}}
\end{figure}

%\begin{figure}
%\epsscale{.60}
%\includegraphics[scale=0.32,width=7cm, height=8cm,angle=-90]{fig7.eps}
%\caption{As for Fig~6, but a close-up of the X-ray data. \label{fig:nustarX}}
%\end{figure}

{\sc optxagnf} assumes the two Comptonization regions to be geometrically co-spatial and that the maximum temperature of the disc is that at ${\rm R_{cor}}$ (Table~1).  The model  is angle-averaged,  corresponding to an inclination angle of 60 degrees.  Input parameters for the model are the nuclear black hole mass and the distance to the source, the black hole spin, the bolometric Eddington ratio $\frac{L}{L_{Edd}}$, ${\rm R_{cor}}$ and the outer disc radius, the electron temperature and optical depth of the low energy electron population, the power law index of the high energy emission and the fraction of the power below the coronal radius which is emitted in the hard Comptonization component, $f_{PL}$.  The model normalization is frozen at 1.0 in the fit, as the model flux is determined by the combination of black hole mass, spin, Eddington ratio and the galaxy distance. Mass estimates for this source range from $1.3 \times 10^8 {\rm M_{\odot}}$ \citep{pounds04b} to 
 $3.8 \times 10^8 {\rm M_{\odot}}$ \citep{oneill05a}.  We have adopted $3.8 \times 10^8 {\rm M_{\odot}}$ for use in the {\sc optxagnf} fits, and this has a corresponding Eddington luminosity $L_{Edd}=4.6 \times 10^{46}$ erg s$^{-1}$.  
 We take the co-moving radial distance of the source to be 418 Mpc 
  \footnote[5]{ned.ipac.caltech.edu}. The electron temperature was limited to be $< 1$ keV for the soft Comptonizing region (to maintain a distinction between the soft and hard Comptonizing regions in the fit and to restrict the fit to a meaningful parameter regime). During fitting we found that kT hit  the limit of 1.0 keV and so it subsequently was fixed at that value, which may be typical for AGN. For comparison, such models have recently been applied successfully to Ark~120, where \citet{porquet17a} find a good fit with temperature $kT=0.5 $ keV and optical depth $\tau \sim 9$ to explain the soft excess \citep[also see][and references within]{rozanska15a}.  

The black hole spin is degenerate with other parameters,  and so we restricted our investigation of spin to a comparison of fits with spin set at the maximum and minimum values of a=0.998 and  a=0.0, respectively. The fit with spin=0.998 was favored at $>99.99\%$ confidence for all variations of fit that are presented in this paper, therefore only the maximal spin fits are tabulated here (Tables~1, 2, 3). However, we caution that while our model constraints offer an interesting case study, we cannot rule out other solutions that combine different model assumptions with a low black hole spin.   Other parameter degeneracies exist. For AGN soft excesses in general, the temperature  of the soft X-ray emitting plasma  parametrized as $kT$, is degenerate with the optical depth, $\tau$. So if  $kT$ is increased in the model then  $\tau$ will decrease in the fit to compensate (and vice versa), because  overall, the luminosity or spectral shape will remain approximately the same.    We note that some of the other degeneracies of the {\sc optxagnf} model are discussed by \citet{thomas16a}. Those authors discuss the practical application of the {\sc optxagnf} model, noting the complete degeneracy between the black hole mass and  luminosity and they discuss how the spectral shape of the AGN SED can be described by three non-degenerate parameters, the  energy of the peak of the accretion disk emission, the photon index of the non-thermal emission and the fraction of the total flux that is emitted in the non-thermal component.  

To account for the Fe K line we again included a reflection component from the 
{\sc xillver} model , with the assumptions generally set as in the reflection fits of Section 3.4.
%except that in this fit the ionization of the reflector was tied to that of the higher column absorber zone (from which we would expect to see significant scattered emission). 
The high energy cut-off of the reflector was  assumed to be 100 keV for consistency with the rollover built into the {\sc optxagnf} model. Importantly, here we did not convolve the reflector with any gravitational blurring, given both the narrow K$\alpha$ line width and flat emissivity profile found previously. The soft line emission table was also added to the model, as previously described.

\begin{figure}
%\epsscale{.90}
\includegraphics[scale=0.42,width=6cm, height=8cm,angle=-90]{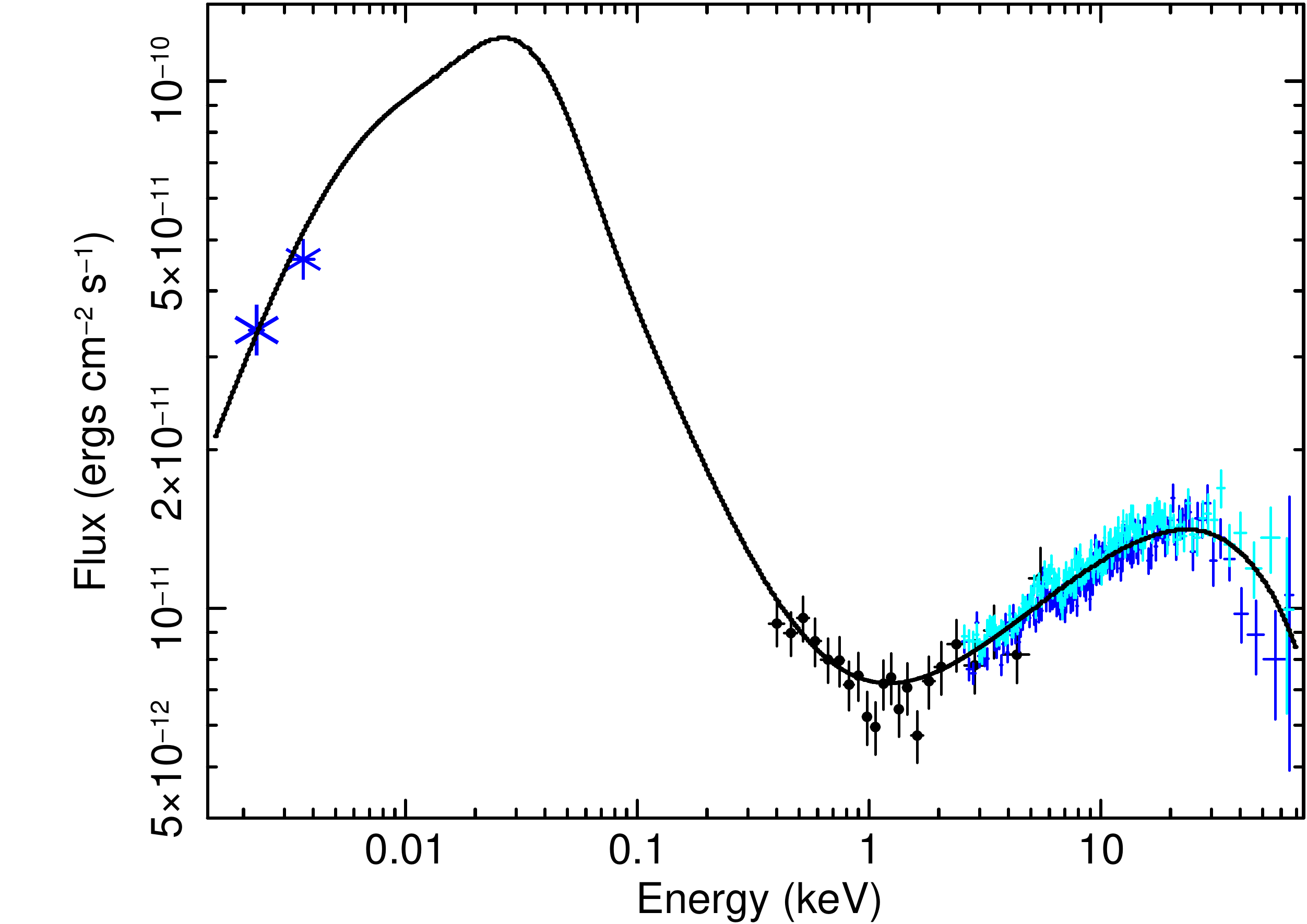}
%\plotone{fig6.eps}
\caption{The fitted {\sc optxagnf} continuum model to 2015 {\it NuSTAR}  and {\it Swift}  data (Table~3, column 4, colours as for Fig.~1). Galactic and intrinsic absorption and host reddening have been removed to show the intrinsic disc plus coronal emission 
from \textsc{optxagnf}. Overlaid on this are the X-ray data points, corrected for absorption (Galactic and intrinsic) and the UVOT U and V  points corrected for Galactic reddening. The UVOT U point was fitted to the data, while the V point (far left) was not included in the fit (because of a concern about its lack of simultaneity, see text for details), but is overlaid for comparison. 
For visual clarity,  the {\sc xstar} emission and {\sc xillver} model components are not shown as these make only a small energetic contribution to the broad emission). \label{fig:model}} 
\end{figure}

Fitting this {\sc optxagnf} plus  {\sc xillver} model to the simultaneous {\it NuSTAR} and {\it Swift} data from 2015 we found that 
even upon the inclusion of ionized reflection in the model, the overall \textsc{optxagnf} fit over-predicts the {\it NuSTAR} flux in the hard X-ray band above 10 keV, while pronounced curvature is still present and the fit is still poor ($\chi^2=1117/953\, d.o.f.$, see Figure~~\ref{fig:underpred}). This is due to the fact that continuum fit is driven by the relatively hard portion of the SED between 2-10 keV (as shown in Figure 1), which requires a relatively hard photon index of $\Gamma=1.8$, while the spectral slope in the hard band above 10 keV is much steeper ($\Gamma>2$). The reflection component in the model only contributes towards a maximum of 10\% of the total flux in the hard band and increasing its strength further leads to a substantially worse fit, as the hard flux then becomes even more discrepant with what is predicted by the model. Thus the broad-band spectrum of 1H\,0419-577 is not likely to be dominated by a strong reflection component, given both the weak Compton hump that is present and the lack of any strong iron K$\alpha$ emission, as described earlier.  
Of course, the reflection component that we use here can arise from any Compton thick gas, it is not necessarily from the accretion disc and indeed, we later suggest that it may arise from clumps of material associated with the thick zone of absorption, seen out of the line-of-sight.

This suggests that either the intrinsic continuum is more complex in form and/or a substantial column of line of sight material is present which modifies the continuum, producing the hard spectral shape between 2-10 keV, while reproducing the underlying steep UV to soft X-ray and also hard X-ray continuum which is apparent in the overall SED shape. 
To test the former, we modified the \textsc{optxagnf} continuum to include an additional ad-hoc exponential cut-off in the {\it NuSTAR} bandpass below 100\,keV, 
while the cut-off value for the reflection model was also tied to that of the illuminating continuum for consistency. While this did result in a good statistical description of the {\it Swift} plus {\it NuSTAR} data (954/952), the resulting continuum shape now
has an even harder photon index of $\Gamma=1.65\pm0.03$ with a very low value for the exponential cut-off of $E_{\rm cut} = 52\pm9$\,keV. Such an extreme hard X-ray continuum shape is flatter than that produced by standard  Comptonization models, although a flat spectrum is predicted for  coronae that have a significant bulk velocity \citep{sobolewska09a}.

Investigating this further (and dropping the additional model cut-off), a single layer of ionized absorption was added to the model to modify the \textsc{optxagnf} continuum,  using the \textsc{zxipcf} model \citep{reeves08a}.  Complex absorption is expected in this source, as a multi-zoned X-ray absorber is seen in X-ray grating data  \citep{digesu13a}, and has been invoked to explain  the X-ray spectral variability observed during the {\it XMM} observations \citep{pounds04a,pounds04b,turner09a}.  The existence of a significant outflowing X-ray absorber is supported by the unambiguous evidence for a disc wind, seen in UV and X-ray spectra of this AGN \citep{dunn07a,digesu13a}. 

We allow the column density, ionization and covering fraction of the absorber to vary, this model represents the clumpy absorber favored by recent observations (see Introduction). 
 {\sc zxipcf}  assumes  a $\Gamma=2.2$ illuminating spectral form over 1-1000 Ryd.  The intrinsic continuum (underlying the reprocessor) appears to be close to this photon index, although, of course,  the observed continuum is often much harder owing to the effects of reprocessing. As the absorber is likely illuminated predominantly by the intrinsic continuum, the assumed SED for {\sc zxipcf} is reasonable for application to this source. 

For consistency, the ionization parameter of the reflector was tied to that of the absorber, assuming they arise from the same material. This yielded a very good fit ($\chi^2=942/952\, d.o.f.$), with values of $N_{\rm H} = 3.1\pm0.3 \times 10^{23}$\,cm$^{-2}$, $\log\xi=0.9\pm0.3$ erg\, cm\, s$^{-1}$ and $f_{\rm cov}=0.37\pm0.03$ for the absorber. Now the photon index is steeper, with $\Gamma=2.0$ and that, plus the effect of the continuum curvature imparted by the absorber above 2\,keV combined with the (modest) $\sim 100$\,keV rollover intrinsic to the {\sc optxagnf} model, is able to reproduce the shape of the hard X-ray spectrum seen in 
the {\it NuSTAR} data. For this model the absorption correction is about 60\% in the 2-10 keV band, yielding an absorption-corrected luminosity $L_{2-10, int} = 6.7 \times 10^{44}\, {\rm erg\,s^{-1}}$.

\subsection{Spectral Variability}
 
Any acceptable model must also account for the spectral variability exhibited between epochs.  First, we consider all the broad-band X-ray data by adding existing 2007 and 2010  {\it Suzaku} datasets  to the fit, linking parameters between the epochs and then allowing variations of individual parameters, in turn, to account for the spectral and flux variability.  We tried several variations of this fit, as follows:

\begin{enumerate}

\item 
The first model comprised {\sc optxagnf} plus {\sc xillver}, with no absorption zones.  The model parameters  were linked, allowing  only $\frac{L}{L_{Edd}}$ to vary between epochs. 
A poor fit was obtained, with $\chi^2=1862/1319\, d.o.f.$  and a hard photon index, $\Gamma=1.79\pm0.01$,  failing to account for the clear spectral curvature and rollover in the hard band (Figure~~\ref{fig:curvature}, top panel). 

\item 
Taking the model from (i), we allowed more freedom in that by letting $\tau$ vary between epochs, which has a strong effect on the models ability to account for the soft excess. This extra freedom resulted in a better fit, with $\chi^2=1585/1316\, d.o.f.$, however it did not account for the high energy rollover in the data (Figure~~\ref{fig:curvature}, panel 2)

\item 
A single zone of partial covering absorption was added to the joint fit, using {\sc zxipcf} as previously described.  In this fit, $\tau$ was linked rather than allowed to vary between epochs, $\frac{L}{L_{Edd}}$  was allowed to vary between epochs. 
 This improved the fit, giving $\chi^2=1433/1312\, d.o.f$. This somewhat improves the fit to the hard band,  and 
 the underlying continuum steepens to  $\Gamma =1.92\pm0.02$  (Figure~~\ref{fig:curvature}, panel 3)

\item 
A second zone of absorption was added to the model using a second {\sc zxipcf} component.  Application of the ``two-zone'' absorber model to the  2015 {\it NuSTAR} and {\it Swift} data alone, provides a good fit  with $\chi^2=925/945\, dof$  (Table 1, column 3).  
This zone improves the fit further for the {\it NuSTAR/Swift} plus {\it Suzaku} data, giving $\chi^2=1291/1309\, 
d.o.f.$ and this provides the first acceptable parameterization of the hard X-ray data (Figure~\ref{fig:curvature}, bottom panel). The underlying photon index has now steepened to $\Gamma=2.00\pm0.03$ (Table~1, column 4). 

\end{enumerate}

In conclusion,  the broadband X-ray spectrum revealed by {\it NuSTAR} and {\it Suzaku}  requires two absorption zones, one of which must be a high column of gas, for the spectral curvature to be modeled correctly. With that in place, the continuum index is required to be steeper, and together these and the intrinsic rollover present in the 
{\sc optxagnf} model allow a good match between data and the model  above 20 keV. Detailed interpretation of the soft band changes is  more ambiguous. In the absence of  grating data we cannot distinguish a variable soft excess (that we have parameterised here by varying $\tau$)  or a variable soft-band absorber \citep[cf][]{digesu14a}. 

\subsection{The two-zone absorber model applied across all flux states}

 Having established a need for two absorber zones modifying the {\sc optxagnf} continuum, we now consider testing that model across all flux states. We note that between  the 2007 and 2010 {\it Suzaku} observations there is no significant difference in spectral shape, although there is a 'colourless' component of variability (i.e. a simple flux change) between those epochs. That flux change cannot be ascribed to changes in the absorbing gas zones in the model, but is likely a change in the underlying continuum level, therefore model parameters (Table~1) were linked to implement that.  
In {\sc optxagnf} a simple flux change can be described in a number of ways, allowing the Eddington ratio, Comptonized fraction or coronal radius to vary. As there is degeneracy in the continuum model, we selected to parametrize this long-timescale flux change by  
allowing only the Eddington ratio to vary between those two epochs (i.e. this is not a unique solution for the source). 

Table~1 shows the fits,  with the parameters linked between the epochs, except for absorber covering fractions and the Eddington ratio (Table 1, columns 4 - 6). 
%Specifically, the coronal radius, photon index, fraction of power in hard component, reflection and soft emission line fluxes were linked for all datasets (as noted previously, the soft lines are modeled separate to the reflection, as \citealt{digesu13a} show that these lines likely arise on the pc scale. 
The ionization of the reflector was linked to the highest column absorber (zone 1 in Table~1).  Zone 1 is of sufficiently high column density that it may produce 
the weak observed scattered spectral component. However we cannot rule out a contribution to the scattered spectral component from other, higher column-density gas. 
The column densities and ionization states of the two absorbers were linked for all datasets, allowing only the 
covering fraction to vary.  
For the two  {\it Suzaku} epochs  we linked the absorber parameters  (owing to the lack of any change in spectral shape between 2007 and 2010).   Obviously, the lack of high quality 
{\it NuSTAR} data for all epochs limits the test of parameter variability across the different flux states, and limits us to a parameterization, rather than a unique solution to the problem.
This joint fit yielded 
$\chi^2=1432/1312\, dof$, with the best spectrum shown over the UV to hard X-ray range in Figure ~\ref{fig:nustar}. 
%and a close-up of the X-ray band in Figure ~\ref{fig:nustarX}. 
The model is also shown in Figure ~\ref{fig:model},  corrected for absorption to show the intrinsic form.

\begin{figure}
%\epsscale{1.0}
\hspace{-2mm}
\includegraphics[scale=0.4,width=8cm, height=8cm,angle=0]{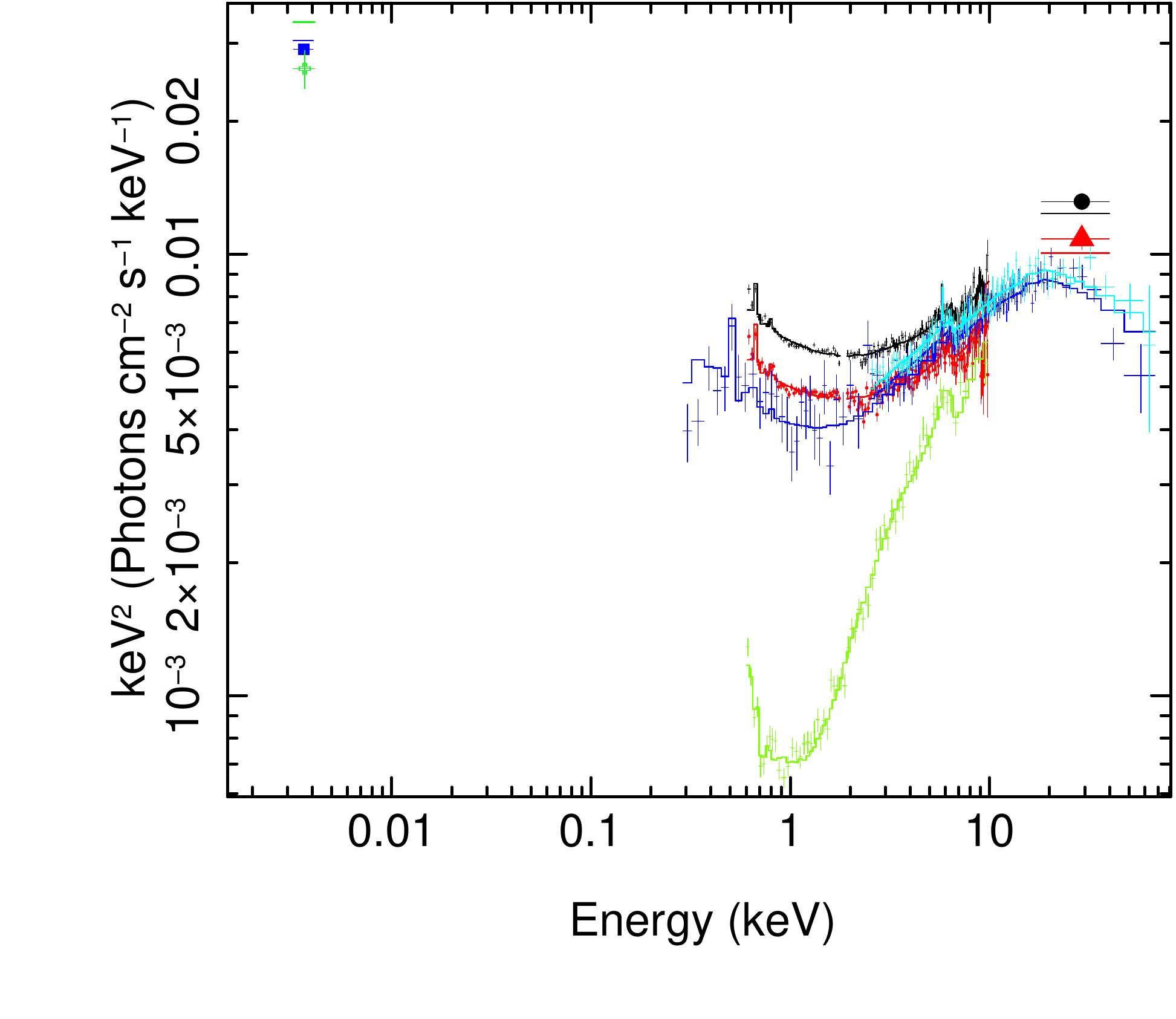}
\caption{The {\it NuSTAR} (blue and aqua) plus simultaneous Swift XRT and UVOT data (blue) compared to the flux states sampled by previous 
{\it XMM} (pn and OM U-band as green, 2002) and {\it Suzaku} (black is 2007, red is 2010) observations.  The solid line is the best fitting model corresponding to Table 1. Note that the low flux {\it XMM} epoch is much more heavily absorbed \label{fig:states}}
\end{figure}

In order to establish whether our general model is capable of accounting for all flux states - in addition to explaining the  {\it NuSTAR/Swift} data and the {\it Suzaku} data, the model should be able to extrapolate to any extreme flux states exhibited by the target. Therefore,  we 
 added to our consideration the extreme low flux X-ray state observed using {\it XMM} \citep{pounds04a,digesu14a}. Previous work on those {\it XMM} data from the low state  shows  clear evidence for a  variable soft X-ray absorber \citep{pounds04b, digesu14a},  and this historical knowledge adds strong support for the  2-zone absorption model to account for the variability over all epochs.   The inclusion of the {\it XMM} data 
produces a fit with $\chi^2=2296/1643\, d.o.f$, providing a poor match to the shape of the lowest  flux state from 2002.  
Considering the evidence to date \citep[e.g.][]{pounds04b} for changes in the opacity of the absorber, we freed the absorber parameters in turn for the {\it XMM} epoch, and refit. 
Freeing the column and ionization of both absorber layers for 2002 data (as well as the covering fraction) produced a significant improvement over the simpler alternatives. With the absorber zones completely free for 2002 data (Table~1 column 7), we achieved  a joint fit statistic $\chi^2=1857/1634\, d.o.f.$ (Figure ~\ref{fig:states}). Indeed it is apparent from the figure that the {\it XMM-Newton} low state spectrum has a substantially harder spectrum, with pronounced absorption seen below 10\,keV. Much of the absorption 
change is driven by an increase in absorber covering for the two zones (see Table\,1). 

%For the low {\it XMM-Newton} epoch the OM U-band (Fig 9) data lies $\sim 2\sigma$ below the corresponding {\it Swift} point. 
%This may either suggest additional long term opt/UV variability in the disc emission or a small change in 
%intrinsic reddening (equivalent to $N_H \sim 4 \times 10^{20} {\rm cm^{-2}} $) associated with increased absorption towards to the source. Further monitoring with {\it Swift} %would be needed to study any long term connection between the UV and X-ray variability.

\begin{table*}
	\centering
	\caption{Variable Eddington Ratio with Partial Covering}
	\label{tab:table}
	\begin{tabular}{cc|c|ccc|c} 
		\hline
		Component & Parameter & 2015$^1$ & 2015 & 2010 & 2007 & 2002 \\
		\hline

		& & NuSTAR/Swift & NuSTAR/Swift  & Suzaku & Suzaku & XMM (Low)  \\
 \hline
 &   L$_{2-10}^2$ & 4.12 & 4.12 & 3.84 & 4.77 &  2.31  \\
& & & & &  & \\
  & L$_{2-10}^3$ & 6.66 &  6.66 & 6.79 & 8.43 &  6.35  \\
 \hline
& & & & &  &  \\
{\sc optxagnf}  & $\Gamma$  	& 2.04$\pm0.03$  & $1.99\pm0.03$ 	& 1.99$^t$ 	& 1.99$^t$ &  1.99$^t$   \\
 & & & & &  &  \\
 	& log\, $\frac{L}{L_{Edd}}^4$ 	&  $ -0.45^{+0.16}_{-0.26}$ & -0.45$^{+0.08}_{-0.13}$	& -0.39$^{+0.07}_{-0.12}$		&  -0.37$^{+0.06}_{-0.09}$    &    $-0.42^{+0.22}_{-0.23}$  \\
 & & & & &  &  \\
 	& ${\rm R_{cor}}$ 	& $15.9^{+47.0}_{-11.1}$ 		& 16.4$^{+50.0}_{-12.2}$ 	& $16.4^t$ 	& $16.4^t$ 	  & $16.4^t$ \\
 & & & & & &   \\
  	& $\tau$ 			& $3.10^{+0.57}_{-1.94}$ 	& 5.47$^{+0.29}_{-1.07}$ 	& 5.47$^t$ 	& 5.47$^t$ 	  & 5.47$^t$   \\
  & & & & & &  \\
   	& $f_{pl}^3$ 	& $0.14^{+0.28}_{-0.24}$		& $0.14^{+0.09}_{-0.08}$ 	& 0.14$^t$ 	&  0.14$^t$ 	   &  0.14$^t$    \\
 & & & & &  &  \\
% {\sc xillver} & log $\xi$ 			& 0.51$\pm0.05$ & $0.51^t$ & $0.51^t$ & $0.51^t$ & $0.51^t$ & $0.51^t$   \\
%& & & & &  & & \\
 {\sc zxipcf$_1^5$} 			& $N_H$ 			& $3.77^{+1.29}_{-0.60}$ & $3.72^{+0.46}_{-0.51}$ & 3.72$^t$ 	& 3.72$^t$  &    1.79$^{+1.29}_{-1.11}$ \\
 & & & & &  & \\
  	& log $\xi$ 		& $0.43^{+0.83}_{-0.30}$	& 0.66$^{+0.32}_{-0.29}$ 	& 0.66$^t$ 	& 0.66$^t$ 	 & -0.37$^{+0.20}_{-0.15}$ \\
  & & & & &  & \\
   	&  $f_{cov}^56$  			& $0.34^{+0.11}_{-0.20}$  &  0.42$^{+0.03}_{-0.05}$ 			& 0.34$\pm0.03$ 		& 0.34$^t$		 &  0.50$^{+0.12}_{-0.09}$ \\
 & & & & &  &  \\
 {\sc zxipcf$_2^5$} 	& $N_H$ 		& $0.30\pm0.55$ 			& 0.17$^{+0.11}_{-0.09}$		& 0.17$^t$ 	& 0.17$^t$ &  0.65$^{+0.50}_{-0.42}$ 
 \\
  & & & & &  & \\
  	& log $\xi$ 		& $ <0.15$	& 0.00$\pm0.50$ 	& 0.00$^t$ 	& 0.00$^t$  	 &    -1.26$^{+0.33}_{-0.87}$ \\
 & & & & &  & \\
  	& $f_{cov}^56$ 			&  $0.34^{+0.18}_{-0.06}$  & 0.57$^{+0.04}_{-0.06}$			&  0.31$\pm0.07$		& $0.31^t$		&   $0.88^{+0.08}_{-0.07}$  \\
  \hline
  $\chi^2/d.o.f.$ & & 925/945 & \multicolumn{3}{c|}{1432/1312} & 1857/1634 \\

		\hline
	\end{tabular}
	\vspace{-0.7cm}
\tablecomments{A column of neutral gas covered all components, fixed at the Galactic 
value ${\rm N_H}=1.26 \times 10^{20}{\rm cm^{-2}} $. 
$^t$ indicates that a parameter was tied to the other epochs during fitting. $p$ indicates the parameter pegged at one of the limits. 
Errors are calculated at 90\% confidence. Reflection was modeled using {\sc xillver}. See text for model details. \\
 $^1$ The 2015 fit was repeated allowing the column densities and ionization parameters to be free of constraint from the other epochs \\
$^2$ The observed 2-10 keV luminosity in units of $10^{44}$ erg\, s$^{-1}$ \\
$^3$ The absorption-corrected 2-10 keV luminosity in units of $10^{44}$ erg\,  s$^{-1}$ \\
$^4$ The units for log $\xi$ are erg cm\,s$^{-1}$ \\
$^5$ Column density in units of 10$^{23} {\rm atom\, cm^{-2}}$ \\
$^6$ Percentage Covering \\
 }
\end{table*}

\section{Discussion}
 
 We have analyzed new data from simultaneous {\it NuSTAR} and {\it Swift} observations of 1H 0419-577 during 2015, in conjunction with previous 
{\it Suzaku} observations from 2007 and 2010 and an {\it XMM} observation \citep{pounds04a,digesu14a} that sampled the extreme low X-ray flux state exhibited during 2002. 
Consideration of the new {\it NuSTAR} data appear to rule out a contribution from strong disc reflection, both due to the shape of the high energy spectrum above 10 keV and the lack of a strong broad Fe  $K\alpha$ line.

Spectral analysis of the {\it NuSTAR}, {\it Swift} and {\it Suzaku} data shows that the source can be fit with a multi-temperature accretion disc, shrouded with a complex clumpy absorber. This model includes both absorption effects and  a scattered light component and is consistent with some of the signatures of variability observed in this target. Adequate parameterization of the very lowest flux state requires changes in the absorber column and ionization state. The source shows a marked turnover at $\sim 20$ keV, that is adequately parameterized by the combined effects of the rollover at 100 keV in {\sc optxagnf}, the effects of the reprocessor and the $\Gamma \sim 2$ index recovered when all reprocessing effects are modeled. 
We cannot rule out the presence of a cut-off in the primary continuum form associated with the high energy corona. 
Application of this model has allowed us to  isolate a 
colourless component of variability effective over timescales of years, that is consistent with changes in the accretion flow on small radial scales in the inner disc.

\subsection{The XUV continuum}
We have analyzed new data from simultaneous {\it NuSTAR} and {\it Swift} observations of 1H 0419-577 during 2015, finding a remarkably low turn-over energy in the hard spectrum. 
 As detailed in \S 3.3, in the context of pure Comptonizing corona models, this marked turnover would imply an electron temperature,   $kT \sim 15$ keV, for the accretion disc, similar to the value reported for Ark~564 \citep{kara16a}, and lower than that reported for  GRS1734-292 \citep{tortosa17a} (also see 
 \citealt{tortosa18a}). 

 In order to solidify our identification of the most compelling modeling solutions, we consider further the implications of the Comptonization model presented in \S 3.3, and in particular the possibility of a low temperature corona  in 1H 0419-577.  Following \citet{fabian15a} (and also \citealt{ dove87a}, \citet{svensson87a}, \citet{lightman87a} ) and taking the coronal radius as $R=R_{cor} = 15 r_g$ and the fitted Eddington ratio $\frac{L}{L_{Edd}} \sim 0.4$ we estimated the source compactness as $l \sim 600$.   Assuming kT=15 keV (Section 3.3) 
 %(estimated as 
%$\frac{E_{cut}}{2}$  from the fitted $E_{cut}=56$ keV \citealt[e.g.][]{petrucci01a, fabian15a})} 
we calculated the source temperature $\Theta=\frac{kT}{m_ec^2} =$0.03. 
 Comparing this with the distribution  of \citet[][their Figure~2]{fabian15a} shows 1H 0419-577 to be outside of the range observed for  AGN in the \citet{fabian15a} sample.   Thus, it appears that `pure' Comptonization models may not yield meaningful results for the cut-off: this led us to try alternative models where at least some of the high energy curvature can be  produced by a higher column partial-covering absorber.

Note that in contrast to 1H\,0419-577, the low black hole mass  of the Narrow Line Seyfert 1 Ark 564 has both a very soft SED, 
which peaks in the EUV and soft X-rays \citep{romano04a} and a steep X-ray photon index, 
where $\Gamma\sim2.6$ \citep{giustini15a}. This might result in substantial Compton cooling of the 
high energy electron pair population, leading to an unusually cool coronal temperature. Furthermore, as noted by \citet{kara16a}, the presence of a hybrid (thermal/non-thermal) plasma \citep{zdziarski93a,fabian17a} could then still maintain a pair-dominated plasma, despite the low electron temperature. On the other hand, in 1H\,0419-577 the black hole mass is likely two orders of magnitude higher than in Ark\,564, leading to an SED peaked in the optical/UV, while the apparently harder photon index (of $\Gamma\sim1.7$) below 10\,keV, (if the effect of absorption is not accounted for in the continuum modeling) may not be able to easily cool the high energy electron population. Thus at least at first sight, 
1H\,0419-577 might appear to be a less plausible AGN for which such a low temperature plasma could 
be maintained.
 Indeed in 1H 0419-577, the cut-off energy and subsequent coronal temperature depends on the continuum model adopted. In the reflection fits (section 3.4), the cut-off energy increases to 60 keV. Furthermore, when any partially covering absorption is accounted for (section 3.5), then the data are consistent with the 100 keV coronal temperature assumed in the {\sc optxagnf} model.
 
We have found that the broadband XUV continuum, sampled using simultaneous  {\it Swift} and {\it NuSTAR} data, can be modeled using a colour-corrected accretion disc while hard X-rays arise via Compton up-scattering from a putative corona \citep{done12a}. The observation of a colourless component of flux change, most evident between 2007 and 2010, indicates the need for continuum variability in addition to the absorber variations.  
Recent work by \citet{digesu14a} presented an alternative, but phenomenologically similar model for the UV to X-ray continuum, based upon contemporaneous {\it XMM} and {\it HST} observations of 1H 0419-577: their parameterization was based around a nested double-Comptonization model.  Consistent with previous modeling, we find the continuum to be modified by reprocessing in two layers of partial-covering gas, including a Compton-thick layer that is consistent with producing the weak Fe K$\alpha$ line observed.  This X-ray absorber imprints a spectral and time variability signature on the data in addition to that imposed by the continuum variation. 

%As noted previously, the mass of the black hole in this source is estimated to be $1.3 -3.8 \times 10^8 {\rm M_{\odot}}$. 
%The 8-1000 $\mu$ luminosity can be measured from the 24 $\mu$ Multiband Imaging Photometer measurement from {\it Spitzer}, and applying the correction factors detailed by \citet{wuyts08a} we find that to be $\sim2.6 \times 10^{45}$ erg s$^{-1}$.  The target is not detected in the {\it Spitzer} 100 $\mu$ filter , nor by {\it Akari} at 90 $\mu$, so we conclude that the bolometric luminosity,  $L_{bol}$, can be estimated from the sum of observed infra-red and XUV luminosities. From the XUV fit the  luminosity over 2 eV - 100 keV is  $L_{XUV} \sim 5.8 \times 10^{45}$ erg s$^{-1}$, giving 
%{\bf $L_{bol} \sim 8.4 \times 10^{45}$ erg s$^{-1}$ } and an expected Eddington ratio $\sim 0.2$. 
The Eddington ratio derived  from the fit to {\sc optxagnf} is  $\sim 0.4$ (Table~1) and this 
 can be related to the viscous timescale. 
The viscous timescale is a characteristic timescale of the mass flow, i.e.  the ratio of the radius of the perturbation in flow to the radial velocity.  To achieve observable modulation of the X-ray continuum flux from changes in the Eddington ratio, viscosity changes or accretion rate changes have to propagate inwards by a significant radial distance and cause  
fluctuations of mass accretion rate in a region of significant energy release. This is only possible for fluctuations on timescales at or exceeding the viscous timescale at the radius where fluctuations are manifested in the accretion flow. 
We consider the simple colourless  flux variability observed between 2007 and 2010. For an AGN with a supermassive black hole of mass $10^8 M_{\odot}$, operating at an Eddington ratio of 0.4,  interpretation of a $\sim 900$ day variability timescale  as a 
viscous timescale, yields a radius $\sim 10r_g$  for the fluctuations in the accretion flow \citep{czerny04b}.  
%While interesting, we would require simultaneous UV/optical photometry for all epochs, to establish that the bolometric luminosity had varied.  
%(An alternative model, where changes in the inner flow are parameterized directly as variations in $R_{cor}$, yield consistent estimates for variations in the flow between $6 - 13\, r_g$).

%The existing data are also consistent with changes in the X-ray to UV flux ratio that may originate from changes in the corona.  In this parametrization, we see a change in %the Comptonized fraction from 10\% to 23\% of the disc emission across available data. 
%Changes may occur in the Comptonized fraction because of variations in coronal conditions. The model implemented here assumes a constant optical depth for the corona, %restricting the possibilities. However, magnetic reconnections could dissipate power into the corona \citep[e.g.][]{blandford82a}
%and cause variations of the X-ray emission  \citep[down to timescales of days,][]{haardt97a} even under conditions of a constant accretion rate.  An alternative interpretation %of changes in $f_{PL}$ is changes in the extent of the corona, that could affect the Comptonized fraction without necessarily affecting optical depth. 

\subsection{The X-ray absorber}

The absorber complex is very important in shaping the spectral and timing behavior of 1H 0419-577. Previous work by 
\citet{pounds04a}  and \citet{digesu14a}  interpreted 
spectral variability below 10 keV as changes in the X-ray absorber, finding that when 
the source brightened the opacity of this absorber decreased, consistent with an increase in gas  
ionization state.  \citet{turner09b}  found evidence in {\it Suzaku} data for a Compton thick component of absorption, extending the \citet{pounds04a}   and \citet{digesu14a}  model to higher energies.  \citet{turner09b} discussed how the gas may be part of a disc wind, and estimated a 10\% global covering for the Compton-thick zone of gas, based on the Fe K$\alpha$ luminosity, and a wind opening angle $ \simeq 12^{\circ}$.   We have applied a new variation on the X-ray model, based upon new broadband data,  and while this has resulted in different parameter values than those found either by 
\citet{pounds04a} or  \citet{turner09b}, our results are in agreement with those previous frameworks: a multi-zoned absorber is required to explain the range of behavior exhibited by 1H 0419-577 in the X-ray regime.  The spectral variability between epochs could be either explained by variations in the absorber covering fraction and/or column density. However as there is only  one {\it NuSTAR} epoch with high quality hard X-ray data available, for simplicity we chose to tie the  column densities while allowing the covering fraction to vary.  This is not a unique solution. 

There are several compelling lines of evidence that indicate the X-ray absorber to be part of a complex outflow of material: \citet{tombesi11a}  
report Fe {\sc xxvi} $Ly\, \alpha$ and Fe {\sc xxv} absorption from gas outflowing at $\sim 24,000$ km/s, while \citet{digesu13a} 
 detail an absorber that has signatures across the UV and soft X-ray regimes, with several kinematic components evident in the UV data, tracing outflow components over 38 -- 220 km/s (as previously observed in other UV spectra of this AGN, e.g. \citealt{dunn07a}).  
  
 For a black hole of mass  $3.8 \times 10^8 {\rm M_{\odot}}$, $1r_g$ is approximately $10^{14}$ cm.  Assuming the continuum source size to be of order 10 $r_g$, for 1H 0419-577 that is a size of $\sim 10^{15}$\, cm. Under an assumption that the cloud size is at least the size of the continuum source  \citep[similar, for example, to NGC 1365,][]{risaliti07a,braito14a} then for the clouds we have 
   $\Delta\, R \sim 10\, r_g \sim 10^{15} $ cm.  The cloud density, $n_e$, can then be estimated from the column density, $N_H$ in the fit, and the radial extent, using $N_H = n_e \times \Delta\, R$, which yields $n_e \sim 10^9 {\rm cm^{-3}}$ for the thick clouds in zone 1.  The radial distance of the zone can then be derived from using this density estimate and the definition of ionization parameter (see Section 3.3) to give a radial distance  $r_{z1} \sim 10^{18}$ cm  from the illuminating source.  We note that this is simply an order-of-magnitude estimate. 
The scale of the optical broad-line region gas  in this source is $r_{BLR}  \sim  10^{17}$cm  \citep{guainazzi98a} and  so the thick absorbing zone  is consistent with the outer broad line region or the putative clumpy torus.  However, the timescale on which changes in covering fraction are observed in this and other AGN \citep[e.g.][]{pounds04b,risaliti07a,bianchi12a} suggests a location on sub-pc scales to be more plausible than very extended kpc scales (such as that found  for the lower column soft X-ray emitting gas \citealt{digesu13a,digesu17a}).

  \citet{digesu13a} 
conclude that a low-ionization absorbing gas zone measured during the 2010 XMM observations must exist at radii $> 3$ kpc from the nucleus, i.e. on a much larger scale than either of the absorber zones detailed here, possibly associated with the galaxy interstellar medium. 
It is  not clear that the high-column clumps in our model are required to be co-spatial with the lower column gas.  \citet{digesu17a}  find evidence of large-scale, geometrically thin, X-ray line emitting gas in {\it Chandra} imaging data. That gas exists on scales of several kpc from the nucleus. One possible origin they suggest is gas that is shock-heated by a wind, on larger scales. We find the high-column component of the  X-ray gas  to be clumpy and highly variable on timescales of months-years \citep{pounds04a, pounds04b} - this component therefore is likely closer in to the nucleus (perhaps on the pc or sub-pc scale) than the other X-ray absorbing zones. One possible link is that the inner X-ray wind, for instance as is predicted in radiatively-driven wind simulations such as in \citet{sim10b}, is driving the gas out to larger scales and seen in the form of the larger scale outflow in the  {\it Chandra} and {\sc [Oiii]}  images.  It seems likely that the various absorbers \citep{digesu13a} found in this AGN form part of a multi-phase outflow. Some of the broad Fe K $\alpha$ features in AGN  could  arise from the forward-scattered emission from such a wind and  indeed, existing broad Fe K components in a small sample of unobscured AGN have been found to be for consistent with the line profiles predicted by the \citet{sim10b} model \citep{tatum12a}.

\section{Conclusions}

We have analyzed and modeled spectral data from a simultaneous {\it NuSTAR} and {\it Swift} observation of 1H 0419-577 during 2015. The broadband spectral form shows a pronounced turnover around 20 keV. We have shown that the rollover cannot be due only to the presence of a strong Compton hump: instead it is likely due to both the presence of thermal Comptonization producing the high energy spectrum and the modifying effects of absorption on the overall broad spectrum, the latter also producing a steeper-than-observed intrinsic continuum once accounted for in the spectral modeling.

The broadband continuum form can be modeled using a centre-corrected accretion disc  around a maximally spinning black hole. In the context of this model, hard X-rays are produced via Compton up-scattering in a corona having $\tau \sim  5$.
Multiple epochs of X-ray data from {\it Suzaku} and {\it NuSTAR}/{\it Swift} show that flux variations on timescales of years are  consistent with modest changes in the source Eddington ratio, likely attributable to fluctuations in the inner disc flow, around $10\, r_g$.  The models tested here are not unique representations of the variability -   alternatives such as changes  in the Comptonizing corona cannot be ruled out - but our analysis showcases some of the simple 
and interesting possibilities to explain these data.

\section*{Acknowledgements}

TJT acknowledges NASA grant NNH13CH63C.  JNR acknowledges NASA grant NNX15AV18G. 
We are  grateful to the {\it NuSTAR} operations team  for performing  this observation and providing 
software and calibration for the data analysis.  We thank the anonymous referee for comments which improved this manuscript. 
This research has also made use of data obtained from the High Energy Astrophysics Science 
Archive Research Center (HEASARC), provided by NASA's Goddard Space Flight Center and from the NASA/IPAC Extragalactic Database (NED), which is operated by the Jet Propulsion Laboratory, California Institute of Technology, under contract with NASA.

%%%%%%%%%%%%%%%%%%%%%%%%%%%%%%%%%%%%%%%%%%%%%%%%%%



\bibliographystyle{mnras}
\bibliography{xray_oct2016} % if your bibtex file is called example.bib


%%%%%%%%%%%%%%%%% APPENDICES %%%%%%%%%%%%%%%%%%%%%

%\appendix


If you want to present additional material which would interrupt the flow of the main paper,
it can be placed in an Appendix which appears after the list of references.

%%%%%%%%%%%%%%%%%%%%%%%%%%%%%%%%%%%%%%%%%%%%%%%%%%

% Don't change these lines
\bsp	% typesetting comment
\label{lastpage}
\end{document}